\documentclass[12pt]{article}
\usepackage{latexsym}
\renewcommand{\epsilon}{\varepsilon} 
\begin{document} 
\title{Efficient Quantum Key Distribution Scheme 
 And Proof of Its Unconditional Security} 
\author{Hoi-Kwong Lo\footnote{email: hklo@comm.utoronto.ca} \\  Department of Electrical and Computer
Engineering; \\
and Department of Physics \\  University of Toronto \\ 10 King's College
Road, Toronto, ON Canada M5S 3G4
\\  H. F. Chau\footnote{email: hfchau@hkusua.hku.hk} \\  Department of Physics, University of Hong Kong, \\  Pokfulam Road, Hong Kong \\  and \\  M. Ardehali  } 
\date{\today} 
\maketitle 
\begin{abstract} 
 We devise a simple modification that essentially doubles the 
 efficiency of the BB84 quantum key distribution scheme proposed 
 by Bennett and Brassard. We also prove the security of our modified 
 scheme 
 against the most general eavesdropping attack that is allowed by the 
 laws 
 of physics. The first major ingredient of our scheme is the 
 assignment of 
 significantly different probabilities 
 to the different polarization bases during both 
 transmission and reception, thus reducing the fraction of 
 discarded data. 
 A second major ingredient of our scheme is 
 a refined analysis of accepted data: We divide the accepted data 
 into various subsets according to the basis employed 
 and estimate an error rate for each subset {\em separately}. 
 We then show that such a refined data analysis guarantees the 
 security of our scheme against 
 the most general eavesdropping strategy, thus generalizing 
 Shor and Preskill's proof 
 of security of BB84 to our new scheme. Up till now, 
 most proposed proofs of security of single-particle type 
 quantum key distribution schemes have relied heavily 
 upon the fact that the bases are chosen uniformly, randomly and 
 independently. Our proof removes this symmetry requirement. 
\end{abstract} 
 
 {\par\medskip\noindent
\begin{minipage}[t]{5in} 
\end{minipage} 
 } 
 
\section{Introduction} 
\label{S:Intro} 
 Since an encryption scheme is only as secure as its key, 
 key distribution is a big problem in 
 conventional cryptography. 
 Public-key based key distribution 
 schemes such as the Diffie-Hellman scheme~\cite{DH} 
 solve the key distribution problem by 
 making computational assumptions such as that 
 the discrete logarithm problem 
 is hard. However, unexpected future advances in algorithms and 
 hardware (e.g., the construction of a quantum computer 
\cite{Shor94,Shor95}) may render many 
 public-key based schemes insecure. Worse still, this would lead 
 to a {\em retroactive\/} total security break with disastrous 
 consequences. This is because an eavesdropper may save a message transmitted
 in the year 2003 and wait for the invention of a new algorithm/hardware
 to decrypt the message decades later.
 A big problem in conventional public-key cryptography is that there 
 is, in 
 principle, nothing to 
 prevent an eavesdropper with infinite computing power 
 from passively monitoring the key distribution 
 channel and thus successfully decoding any subsequent communication.

 Recently, there has been much interest in using quantum mechanics 
 in cryptography. (The subject of quantum cryptography was 
 started by S. Wiesner \cite{Wiesner} in a paper that was written 
 in about 1970 but remained unpublished until 1983. For reviews 
 on the subject, see \cite{sciam,pt,book}.) 
 The aim of quantum cryptography has always been to solve problems 
 that are impossible from the perspective of conventional 
 cryptography. 
 This paper deals with quantum key distribution \cite{BB84,Mor,Ekert} 
 whose 
 goal is to detect eavesdropping using the laws of 
 physics.\footnote{Another class of applications of 
 quantum cryptography has also been proposed \cite{BBCS,BCJL}. 
 Those applications are mainly 
 based on quantum bit commitment and quantum one-out-of-two oblivious  transfer. 
 However, it is now known \cite{Mayers2,LoChau1,LoChau2,Lo} that 
 unconditionally secure quantum bit commitment and 
 unconditionally secure quantum one-out-of-two oblivious 
 transfer are both 
 impossible. Furthermore, other quantum cryptographic schemes such as  a general two-party secure computation have 
 also been shown to be insecure \cite{Lo,LoChau2}. 
 For a review, see \cite{special}.} 
 In quantum mechanics, measurement is not just a passive, external 
 process, 
 but an integral part of the formalism. 
 Indeed, thanks to the quantum no-cloning 
 theorem \cite{Dieks82,WZ82}, passive monitoring of unknown 
 transmitted signals 
 is strictly forbidden in quantum mechanics. 
 Moreover, an eavesdropper who is listening to a channel in an 
 attempt 
 to learn information about quantum states will almost always 
 introduce 
 disturbance in the transmitted quantum signals~\cite{BBM}. 
 Such disturbance can be detected with high probability 
 by the legitimate users. Alice and Bob will use the transmitted 
 signals as a key for subsequent communications only when the 
 security of quantum signals is established 
 (from the low value of error rate). 
 
 Although various QKD schemes have been proposed, the 
 best-known one is still perhaps the first QKD scheme 
 proposed by Bennett and Brassard and published in 1984 \cite{BB84}. 
 Their scheme, which is commonly known as the BB84 scheme, will be 
 briefly 
 discussed in Section~\ref{sec:bb84}. Here 
 it suffices to note two of its characteristics. 
 First, in BB84 each of the two users, Alice and Bob, 
 chooses for each photon between two polarization
bases randomly (that is, the choice of basis is a 
 random 
 variable), \emph{uniformly} (that is, with equal probability) and 
 independently. For this reason, 
 half of the times they are using different basis, in which case 
 the data are rejected immediately. Consequently, the efficiency of 
 BB84 is at most 50\%. Second, a simple-minded error analysis is 
 performed 
 in BB84. That is to say, all the accepted data (those that are 
 encoded and 
 decoded in the same basis) are lumped together and a \emph{single} 
 error rate is computed. 
 
 In contrast, in our new scheme Alice and Bob choose between the two 
 bases 
 randomly, independently but \emph{not} uniformly. 
 In other words, the two bases 
 are chosen with \emph{substantially different} probabilities. 
 As Alice and Bob are now much more likely to be using the same 
 basis, the fraction of discarded data is greatly reduced, 
 thus achieving a significant gain in efficiency. 
 In fact, we are going to show 
 in this paper that the efficiency of our scheme can be made 
 asymptotically 
 close to unity. 
 (The so-called orthogonal quantum cryptographic schemes have also 
 been 
 proposed. They use only a single basis of communication and, 
 according to Goldenberg, it is possible to use them to achieve 
 efficiencies 
 greater than $50\%$ \cite{gold,koashi}. Since they are conceptually 
 rather different from what we are proposing, we will not discuss 
 them here.) 
 
 Is the new scheme secure? If a simple-minded error analysis like the 
 one that lumps all accepted data together were employed, 
 an eavesdropper could easily break a scheme by eavesdropping 
 mainly along the predominant basis. 
 To ensure the security of our scheme, it is crucial to 
 employ a refined data analysis. That is to say, the accepted data 
 are further divided into two subsets according to 
 the actual basis used by Alice and Bob and 
 the error rate of each subset is computed \emph{separately}. 
 We will argue in this paper that such a refined error analysis is 
 sufficient in ensuring the security of our improved scheme, 
 against the \emph{most general} type of eavesdropping attack allowed 
 by the laws of quantum physics. This is done by 
 using the technique of Shor and Preskill's proof \cite{shorpre} of 
 security of BB84 --- a proof that built on the earlier work 
 of Lo and Chau \cite{qkd} and of Mayers \cite{mayersqkd}. 
 
 Our scheme is worth studying for several reasons. First, unlike 
 the entanglement-based QKD scheme proposed by Lo and Chau in 
 Ref.~\cite{qkd}, 
 the implementation of our new scheme does \emph{not} require 
 a quantum computer. It only involves the preparation and 
 measurement of single photons as in standard BB84. 
 Second, none of the existing schemes 
 based on non-orthogonal quantum cryptography has an 
 efficiency more than $50\%$. (We shall say a few word on the 
 so-called 
 orthogonal quantum cryptography in Section~\ref{sec:conclusion}.) 
 By showing in this paper that the efficiency of our new scheme can 
 be made asymptotically close to 100\%, we know that QKD can be made 
 arbitrarily 
 efficient. Our idea is rather general and can be applied to improve 
 the efficiency of some other existing single particle based QKD 
 schemes such as the six-state scheme\cite{bruss,six}). 
 Note that the efficiency of quantum cryptography is of 
 practical importance because it may play an important role 
 in deciding the feasibility of practical quantum cryptographic 
 systems in any future application. Third, our scheme is one of the 
 few QKD schemes whose security have been rigorously 
 proven. Finally, all previous proofs of security seem to rely 
 heavily on 
 the fact that the two bases are chosen randomly and uniformly. 
 Our proof shows that such a requirement is redundant. 
 Another advantage of our security proof is that it does not 
 depend on asymptotic argument and hence can be applied 
 readily to realistic situation involving only a relatively small 
 amount of quantum signal transmission. 
 
 The organization of our paper is as follows. The basic features and 
 the requirements of unconditional security will be reviewed in 
 Section~\ref{sec:basic}. In Section~\ref{sec:bb84}, we will review  the BB84 scheme and Shor-Preskill proof for completeness. Readers who  are already familiar with the BB84 scheme and Shor-Preskill proof may  browse through Section~\ref{sec:basic} and skip Section~\ref{sec:bb84}.  An overview of our proof of security of 
 an efficient QKD scheme will be given in Section~\ref{sec:overview}, 
 which is followed by Section~\ref{s:proof} which ties up some loose 
 ends. Finally, we give some concluding remarks in 
 Section~\ref{sec:conclusion}. 
 
\section{Basic Features and Requirements of a Quantum Key 
 Distribution Scheme} 
\label{sec:basic} 
\hspace{2.7ex} 
 
\subsection{basic procedure} 
\label{subsec:procedure} 
 The aim of a QKD scheme is to allow two cooperative participants 
 (commonly 
 known as Alice and Bob) to establish a common secret key in 
 the presence of noise and eavesdropper (commonly known as Eve) 
 by exploiting 
 the laws of quantum physics. More precisely, it is
commonly assumed that Alice and Bob share a small amount of initial
authentication information. The goal is then to expand such
a small amount of authentication information into a long secure key.
In almost all QKD schemes proposed 
 so far, 
 Alice and Bob are assumed to have access to a classical public 
 unjammable channel as well as a quantum noisy 
 insecure channel. 
 That is to say, we assume that everyone, including the 
 eavesdropper 
 Eve, can 
 listen to the conversations but cannot change the message that 
 send through the 
 public classical channel. In practice, an authenticated 
 classical channel should suffice. On the other hand, 
 the transmission of quantum signal can be done 
 through free air \cite{exp,Buttler,free} or optical 
 fibers \cite{hughes00,MZG95,townsend98} in practice. 
 The present state-of-the-art quantum channel for QKD can 
 transmit signals up to a rate of $4\times 10^5$~qubits per 
 second over a 
 distance of about 10~km with an error rate of a few percent 
\cite{Buttler,hughes00,townsend98}.\footnote{In 
 experimental implementations, 
 coherent states with a Poisson distribution in the 
 number of photons are often employed. To achieve unconditional 
 security, it is important that the operational parameters are 
 chosen such that the fraction of multi-photon signals is 
 sufficiently 
 small. This may substantially reduce the key generation 
 rate\cite{ilm}. In the current 
 paper, we restrict our attention to perfect single photon 
 signals as assumed in standard BB84 and various security proofs.} 
 The quantum channel is assumed to be insecure. That is to say 
 that 
 the eavesdropper is free to manipulate the signal transmitted 
 through the 
 quantum channel as long as such manipulation is allowed 
 by the known laws of 
 physics. 
 
 Using the above two channels, procedures in all secure QKD 
 schemes we 
 know of to date can be divided into the following three stages: 
\begin{enumerate} 
\item \emph{Signal Preparation And Transmission Stage}: Alice 
 and Bob 
 separately prepare a number of classical and quantum signals. 
 They 
 may keep some 
 of them private and transmit the rest to the other party using 
 the secure 
 classical and insecure quantum channels. They may iterate the 
 signal 
 preparation and transmission process a few times. 
\item \emph{Signal Quality Check Stage}: Alice and Bob then (use 
 their 
 private information retained in the signal preparation and 
 transmission 
 stage, the secure classical channel and their own quantum 
 measurement apparatus 
 to) test the fidelity of their exchanged quantum signals that 
 have just been 
 transmitted through the insecure and noisy quantum channel. 
 Since a quantum measurement is an irreversible process 
 some quantum signals are consumed in 
 this signal quality check stage. The aim of their test is to 
 estimate the 
 noise and hence the upper bound for the eavesdropping level of 
 the channel from 
 the sample of quantum signals they have measured. In other 
 words, the process 
 is conceptually the same as a typical quantity control test in a 
 production 
 line --- to test the quality of products by means of destructive 
 random 
 sampling tests. Alice and Bob abort and start all over again in 
 case they 
 believe from the result of their tests that the fidelity of the 
 remaining 
 quantum signal is not high enough. Alice and Bob proceed to the 
 final stage 
 only if they believe from the result of 
 their tests that the fidelity of the remaining quantum signal is 
 high. 
\item \emph{Signal Error Correction and Privacy Amplification 
 Stage}: Alice and Bob need to correct errors in 
 their remaining signals. Moreover, they would like to remove any 
 residual information Eve might still have on the signals. In other 
 words, 
 Alice and Bob would like to distill from the remaining 
 untested quantum signals a smaller set of almost perfect 
 signals without being eavesdropped or corrupted by noise. We 
 call this 
 process privacy amplification. Finally, Alice 
 and Bob make use of these distilled signals to generate their 
 secret shared key. 
\end{enumerate} 
 
\subsection{security requirement} 
 
 A QKD scheme is said to be secure if, for any eavesdropping 
 strategy by Eve, either a) it is highly unlikely 
 that the state will pass Alice 
 and Bob's quality check stage or b) with a high 
 probability that Alice 
 and Bob will share the same key, which is essentially random 
 and, furthermore, Eve has a negligible amount of information 
 on their shared key.\footnote{Naively, one might think 
 that the security requirement should simply be: conditional on 
 passing the quality check stage, Eve has a negligible amount of 
 information on the key. However, such a strong security 
 requirement is, in fact, impossible to 
 achieve \cite{mayersqkd,qkd}. The point is that a determined 
 eavesdropper can 
 always replace all the quantum signals from Alice by some specific 
 state prepared by herself. Such a strategy will most likely 
 fail in the quality check. But, if it is lucky enough to pass, 
 then Eve will have perfect information on the key shared 
 by Alice and Bob.}

\section{Bennett and Brassard's Scheme (BB84)} 
\label{sec:bb84} 
 
\subsection{Basic idea of the BB84 scheme} 
\label{subsec:basicidea} 
\hspace{2.7ex} 
 We now briefly review the basic ingredients of the BB84 scheme 
 and the 
 ideas behind its security. Readers who are already familiar with  BB84 and the Shor-Preskill proof may choose to skip this section to  go directly to our biased scheme in Section~4.  In BB84 \cite{BB84}, Alice prepares and transmits to 
 Bob a batch of photons each of which is independently in one of 
 the four 
 possible polarizations: 
 horizontal ($0^\circ$), vertical ($90^\circ$), $45^\circ$ and 
 $135^\circ$. 
 For each photon, Bob randomly picks one of the two (rectilinear 
 or diagonal) 
 bases to perform a measurement. While the measurement outcomes 
 are kept secret by Bob, Alice and Bob publicly compare their 
 bases. They keep only the polarization data that are 
 transmitted and received in the same basis. 
 Notice that, in the absence of noises and 
 eavesdropping interference, those polarization data should 
 agree. 
 This completes the signal preparation and transmission stage of 
 the BB84 
 scheme. 
 We remark that the laws of quantum physics strictly forbid Eve 
 to distinguish between the four possibilities with certainty. This 
 is because 
 the two polarization bases, namely rectilinear and diagonal, are 
 complementary 
 observables and quantum mechanics forbids the simultaneous 
 determination of 
 the eigenvalues of complementary observables.\footnote{Mathematically,
observables in quantum mechanics are represented by Hermitian matrices.
Complementary observables are represented by non-commuting matrices and,
therefore, cannot be simultaneously diagonalized. Consequently, their
simultaneous eigenvectors generally do not exist.} 
 More importantly, any eavesdropping attack will lead to 
 a disagreement in the polarization data between Alice and Bob, 
 which can be 
 detected by them through public classical discussion. 
 More concretely, to test for tampering in the signal quality 
 check stage, 
 Alice and Bob choose a random 
 subset of the transmitted photons and publicly compare their 
 polarization data. If the quantum bit error rate (that is, the 
 fraction of 
 polarization data that disagree) is unreasonably large, they 
 throw away all polarization data and start all over again.
 On the other hand, 
 if the quantum bit error rate is acceptably small, 
 they should then move on to the signal error correction and 
 privacy amplification stage by 
 performing public classical discussion to correct remaining 
 errors. 
 
 Proving security of a QKD scheme turned out to be a very tricky 
 business. The problem is that, in principle, Eve may 
 have a quantum computer. Therefore, she could employ a 
 highly sophisticated eavesdropping attack by entangling all the 
 quantum 
 signals transmitted by Alice. Moreover, she could wait to hear 
 the subsequent classical discussion between Alice and Bob during 
 both 
 the signal quality check and the error correction and privacy 
 amplification stages before making any measurement on her 
 system.\footnote{As demonstrated by the well-known 
 Einstein-Podolsky-Rosen 
 paradox, classical intuitions generally do not apply to quantum 
 mechanics. This is a reason why proving security of QKD is hard.} 
 One class of proofs by Mayers \cite{mayersqkd} and subsequently 
 others \cite{benor,biham} proved the security of the standard BB84 
 directly. Those proofs are relatively complex. 
 Another approach by Lo and Chau \cite{loqkd,qkd} dealt with 
 schemes that are based on quantum error-correcting codes. 
 It has the advantage of being conceptually 
 simpler, but requires a quantum computer to implement. These two 
 classes of proofs have been linked up by the recent seminal work 
 of Shor and Preskill \cite{shorpre}, who provided a simple proof of 
 security of the BB84 scheme. They showed that an eavesdropper is 
 no better off with standard BB84 than a QKD scheme based on 
 a specific class of quantum error-correcting codes. 
 So long as from Eve's view, 
 Alice and Bob {\it could have} performed the key generation by using 
 their quantum computers, one can bound Eve's information on the key. 
 It does not matter that Alice and Bob did not really use 
 quantum computers. 
 
\subsection{entanglement purification} 
 
 To recapitulate Shor and Preskill's proof, we shall first introduce 
 a QKD scheme based on entanglement purification and prove its 
 security. 
 Our discussion in the next few subsections 
 essentially combines those of Shor and 
 Preskill\cite{shorpre} and Gottesman and 
 Preskill\cite{squeeze}.\footnote{There are some subtle 
 differences between the original Shor and Preskill's proof and 
 the one elaborated by Gottesman and Preskill. First, 
 in the original Shor and Preskill's proof, Alice and 
 Bob apply a simple-minded error rate 
 estimation procedure in which they lump all polarization data of 
 their test sample together into a single set and 
 compute a single bit error rate. In contrast, 
 in Gottesman and Preskill's elaboration, 
 Alice and Bob separate the polarization data according to the 
 bases in which they are transmitted and received. 
 The two bit error rates for the rectilinear 
 and diagonal bases are computed separately. In essence, 
 they are employing the refined data analysis idea, which was first 
 presented in a preliminary version of this manuscript \cite{prelim}.  Second, in Gottesman and Preskill's discussion, the final 
 key is generated by measuring along a {\it single} basis, 
 namely the $Z$-basis. (Because of this prescription, they 
 call the error rates of the two bases simply bit-flip 
 and phase errors. To avoid any potential confusion, 
 we will not use their terminology here.) 
 In contrast, in Shor and Preskill's original proof, the final 
 key is generated from polarization data obtained in both bases.} 
 
 Entanglement purification was first proposed by Bennett, DiVincenzo, 
 Smolin and Wootters (BDSW) \cite{BDSW}. 
 Its application to QKD was first proposed 
 by Deutsch {\it et al.} \cite{deutsch}. A convincing 
 proof of security based on entanglement purification 
 was presented by 
 Lo and Chau \cite{qkd}. Finally, Shor and Preskill\cite{shorpre} 
 noted its connection to BB84. 
 
 Suppose two distant observers, Alice and Bob, share $n$ 
 impure EPR pairs. That is to say, some noisy version of the 
 state 
\begin{equation} 
 | \Phi^{(n)} \rangle = | \Phi^+ \rangle^{\otimes n} 
\end{equation} 
 where $  | \Phi^+ \rangle = 
 { 1 \over \sqrt 2}  ( | 00 \rangle + | 11 \rangle ) $. 
 They may wish to distill out a smaller 
 number, say $k$, pairs of perfect EPR pairs, by applying only 
 classical communications and local operations. This 
 process is called entanglement purification \cite{BDSW}. 
 Suppose they succeed in generating 
 $k$ perfect EPR pairs. 
 By measuring the resulting EPR pairs along a common axis, 
 Alice and Bob can obtain a secure $k$-bit key. 
 
 Of course, a quality check stage must be added in QKD 
 to guarantee the likely success of the entanglement purification 
 procedure (for any eavesdropping attack that will pass the 
 quality check stage with a non-negligible probability). 
 A simple quality check procedure is for Alice and Bob to 
 take a random sample of the pairs and measure each of them 
 randomly along either $X$ or $Z$ axis and compute the 
 bit error rate (i.e., the fraction in which the answer 
 differs from what is expected from an EPR pair). 
 Suppose they find the bit error rates for the $X$ 
 and $Z$ bases of the sample to be $p_X$ and $p_Z$ respectively. 
 For a sufficiently large sample size, 
 the properties of the sample provide good 
 approximations to those of the population. Therefore, 
 provided that the entanglement purification protocol 
 that they employ can tolerate slightly more than 
 $p_X$ and $p_Z$ errors in the two bases, we would 
 expect that their QKD scheme is secure. 
 This point will be proven in subsequent discussions in 
 subsection~\ref{subsec:quality}. 
 
 Let us introduce some notations. 
 
 {\it Definition: Pauli operators.} We define a Pauli operator 
 acting on $n$ qubits 
 to be a tensor product of individual qubit operators that are of the 
 form 
 $I = \pmatrix{1 & 0 \cr 0 & 1}$,
 $X = \pmatrix{0 & 1 \cr 1 & 0}$, 
 $Y = \pmatrix{0 & -i \cr i & \ 0}$ and 
 $Z = \pmatrix{1 & \ 0 \cr 0 & -1}$. 
 
 For example, ${\cal P} = X \otimes I \otimes Y \otimes Z$ is a Pauli 
 operator. 
 
 We shall consider entanglement purification protocols that 
 can be conveniently described by 
 {\it stabilizers}\cite{gottesman96,gottesmanthesis}. 
 A stabilizer is an Abelian group whose generators, $M_i$'s, are 
 Pauli operators.

 Consider a fixed but arbitrary $[[n,k,d]]$ stabilizer-based quantum 
 error-correcting code (QECC). The notation $[[n,k,d]]$ means 
 that it encodes $k$ logical qubits into $n$ physical qubits 
 with a minimum distance $d$. 
 As noted in \cite{BDSW}, the encoding and 
 decoding procedure of Alice and Bob can 
 be equivalently described by a set of Pauli operators, 
 $M_i$, with both Alice and Bob measuring the same 
 operator $M_i$. 
 To generate the final key from the encoded qubits, Alice 
 and Bob eventually apply a set of operators, say 
 $\bar{Z}_{a,A}$ and $\bar{Z}_{a,B}$ respectively, for $a= 1, 2, 
\cdots, k$. 
 In Shor and Preskill's proof, all Alice's (Bob's respectively) 
 operators commute with each other. 
 
 If the $n$ EPR pairs were perfect, Alice and Bob would 
 obtain identical outcomes for their measurements, 
 $ M_{i,A}$ and $M_{i,B}$. Moreover, because of 
 the commutability of the operators, those measurements 
 would not disturb the encoded operations, 
 $\bar{Z}_{a,A} \otimes \bar{Z}_{a,B}$, each of which will 
 give $+1$ as its eigenvalue for the state of $n$ perfect 
 EPR pairs. This is because measurements $\bar{Z}_{a,A}$ 
 and $\bar{Z}_{a,B}$ produce the same $+1$ or $-1$ eigenvalues. 
 
 What about $n$ noisy EPR pairs? Suppose Alice and Bob 
 broadcast their measurement outcomes for $ M_{i,A}$ 
 and $M_{i,B}$ respectively. The product of their 
 measurement outcomes of 
 $ M_{i,A}$ and $M_{i,B}$ gives the error syndrome of the 
 state, which is now noisy. Since the original 
 QECC can correct up to $t \equiv \lfloor { d -1 \over 2} 
\rfloor$ errors, intuitively, provided that the number of 
 bit-flip and phase error errors are each less than $t$, 
 Alice and Bob will successfully correct the state to 
 obtain the $k$ encoded EPR pairs. Now, they 
 can measure the encoded operations 
 $\bar{Z}_{a,A} \otimes \bar{Z}_{a,B}$ to obtain 
 a secure $k$-bit key.

\subsection{Reduction to Pauli strategy} 
\label{subsec:quality} 
 
 {\it Definition: Correlated Pauli strategy.} Recall that a Pauli 
 operator acting on $n$ qubits 
 is defined to be a tensor product of individual qubit 
 operators that are of the form $I$, $X$, $Y$ and $Z$. 
 We define a correlated Pauli strategy, $({\cal P}_i , q_i)$, 
 to be one in which 
 Eve applies only Pauli operators. That is to say that 
 Eve applies a Pauli operator ${\cal P}_i$ 
 with a probability $q_i$. 
 
 The argument in the last subsection is precise only for 
 a specific class of eavesdropping strategies, namely 
 the class of correlated Pauli strategies. 
 In this case, the numbers of bit-flip and 
 phase errors are, indeed, well-defined. 
 What about a general eavesdropping attack? 
 In general, Alice and Bob's system is entangled 
 with Eve's system. Does it still make any 
 sense to say that Alice and Bob's system has no more than 
 $t$ bit-flip errors and no more than $t$ phase errors? 
 Surprisingly, it does. 
 Instead of having to consider all possible eavesdropping 
 strategies by Eve, it turns out that it is 
 sufficient to consider the 
 Pauli strategy defined above. 
 In other words, one can 
 assume that Eve has applied some Pauli operators, 
 i.e., tensor products of single-qubit identities 
 and Pauli matrices, on the transmitted signals 
 with some {\it classical} probability distribution. 
 More precisely, it can be shown that the fidelity of the 
 recovered $k$ EPR pairs is at least as big as the 
 probability that i) $t$ or fewer bit-flip errors 
 and ii) $t$ or fewer phase errors {\it would have} been found if 
 a Bell-measurement had been performed on the $n$ pairs.

 Mathematically, the insight can be 
 stated as the following theorem: 
 
 {\bf Theorem~1 (from \cite{squeeze,shorpre,qkd})}: 
 Suppose Alice 
 and Bob share a bipartite state of $n$ pairs of 
 qubits and they execute 
 a stabilizer-based entanglement purification procedure that 
 can be described by the measurement operators, $M_i$, 
 with both Alice and Bob measuring the same 
 $M_i$. 
 Suppose further that the procedure leads to 
 a $[[n,k,d]]$ QECC which 
 corrects $t \equiv \lfloor { d - 1  \over 2} \rfloor $ bit-flip 
 errors and also $t$ phase errors. Then, 
 the fidelity of the recovered state, after error correction, 
 as $k$ EPR pairs 
\begin{equation} 
 F \equiv \langle  \bar{\Phi}^{(k)} | \rho_R |  \bar{\Phi}^{(k)} 
\rangle \geq 
 {\rm Tr} \left( \Pi_S \rho \right) . 
\label{e:reduction} 
\end{equation} 
 Here, $\bar{\Phi}^{(k)}$ is the encoded state of $k$ EPR pairs, 
 $\rho_R $ is the density matrix of the recovered state 
 after quantum error correction, $\rho$ is the 
 density matrix of the $n$ EPR pairs before 
 error correction and $\Pi_S$ represents the projection 
 operator into the Hilbert space, called ${\cal H}_{\rm good}$, 
 which is spanned by Bell pairs states that differ from 
 $n$ EPR pairs in no more than $t$ bit-flip errors and also 
 no more than $t$ phase errors. 
 
 {\bf Proof of Theorem~1}: 
 
 One can regard $\rho$ as 
 the reduced density matrix of some pure state $ | \Psi \rangle_{SE}$ 
 which describes the state of the system, $S$ and an ancilla 
 (the environment, $E$, outside Alice and Bob's control). 
 Now, in the recovery procedure, Alice and Bob couple some 
 auxiliary reservoir, $R$, prepared in some 
 arbitrary initial state, $| 0 \rangle_R$, to the system. 
 Initially, let us decompose the pure state 
 $| \Psi \rangle_{SE} \otimes | 0 \rangle_R $ 
 into a ``good'' component and a ``bad'' component, where 
 the good component is defined as: 
\begin{equation} 
 | \Psi_{good} \rangle = 
 ( \Pi_S  \otimes I_{ER} ) | \Psi \rangle_{SE} \otimes | 0 \rangle_R 
\end{equation} 
 and 
 the bad component is given by: 
\begin{equation} 
 | \Psi_{bad} \rangle  = 
 ( ( I_S - \Pi_S ) \otimes I_{ER} ) 
 | \Psi \rangle_{SE} \otimes | 0 \rangle_R . 
\end{equation} 
 
 Now, the recovery procedure will map the two components, 
 $| \Psi_{good} \rangle $ and $ | \Psi_{bad} \rangle $, unitarily 
 into 
 $| \Psi'_{good} \rangle $ and $ | \Psi'_{bad} \rangle $. 
 Since the recovery procedure works perfectly in the subspace, 
 ${\cal H}_{\rm good}$, we have 
\begin{equation} 
 | \Psi'_{good} \rangle  =  |  \bar{\Phi}^{(k)} \rangle_S 
\otimes | junk  \rangle_{ER}. 
\end{equation} 
 
 Let us consider the norm of the good component: 
\begin{eqnarray} 
\langle  \Psi'_{good} | \Psi'_{good} \rangle & =& 
\langle  \Psi_{good} | \Psi_{good} \rangle \nonumber \\ 
 & =& {\rm Tr} \left( \Pi_S \rho \right) . 
\end{eqnarray} 
 
 Now, the fidelity of the final state as an $k$-EPR pairs is 
 given by: 
\begin{eqnarray} 
 F &=& ~_{SER}\langle \Psi' | \left( |  \bar{\Phi}^{(k)} \rangle_S 
 ~_S \langle \bar{\Phi}^{(k)} | \right) \otimes I_{ER} | \Psi' 
\rangle_{SER} 
\label{e:fidelity1} \\ 
 ~ &=& ~_{SER}\langle \Psi'_{good} | \left( |  \bar{\Phi}^{(k)} 
\rangle_S 
 ~_S \langle \bar{\Phi}^{(k)} | \right) \otimes I_{ER} | 
\Psi'_{good} \rangle_{SER} 
\nonumber \\ 
 ~ & & + ~_{SER}\langle \Psi'_{bad} | \left( |  \bar{\Phi}^{(k)} 
\rangle_S 
 ~_S \langle \bar{\Phi}^{(k)} | \right) \otimes I_{ER} | 
\Psi'_{bad} \rangle_{SER} 
\nonumber \\ 
 ~ & & + ~_{SER}\langle \Psi'_{good} | \left( |  \bar{\Phi}^{(k)} 
\rangle_S 
 ~_S \langle \bar{\Phi}^{(k)} | \right) \otimes I_{ER} | 
\Psi'_{bad} \rangle_{SER} 
\nonumber \\ 
 ~ & & + ~_{SER}\langle \Psi'_{bad} | \left( |  \bar{\Phi}^{(k)} 
\rangle_S 
 ~_S \langle \bar{\Phi}^{(k)} | \right) \otimes I_{ER} | 
\Psi'_{good} \rangle_{SER} 
\label{e:fidelity2} \\ 
 ~ &=& {\rm Tr} \left( \Pi_S \rho \right) \nonumber \\ 
 ~ & & + ~_{SER}\langle \Psi'_{bad} | \left( |  \bar{\Phi}^{(k)} 
\rangle_S 
 ~_S \langle \bar{\Phi}^{(k)} | \right) \otimes I_{ER} | 
\Psi'_{bad} \rangle_{SER} 
\nonumber \\ 
 ~ & & + ~_{SER}\langle \Psi'_{good} | \Psi'_{bad} \rangle_{SER} 
\nonumber \\ 
 ~ & & + ~_{SER}\langle \Psi'_{bad} | \Psi'_{good} \rangle_{SER} 
\label{e:fidelity3} \\ 
 ~ &=& {\rm Tr} \left( \Pi_S \rho \right) \nonumber \\ 
 ~ & & + ~_{SER}\langle \Psi'_{bad} | \left( |  \bar{\Phi}^{(k)} 
\rangle_S 
 ~_S \langle \bar{\Phi}^{(k)} | \right) \otimes I_{ER} | 
\Psi'_{bad} \rangle_{SER} 
\label{e:fidelity4} \\ 
 ~ &\geq & {\rm Tr} \left( \Pi_S \rho \right) 
\label{e:fidelity5} 
\end{eqnarray} 
 where the orthogonality of the states, $  | \Psi'_{good} 
\rangle_{SER}$ 
 and $ | \Psi'_{bad} \rangle_{SER} $, is used in 
 Eq.~(\ref{e:fidelity4}). \hfill Q.E.D.

\subsection{quality check procedure} 
 
 In the last subsection, we showed that, provided that a Bell 
 measurement, if had been performed, would have 
 shown that the numbers of bit-flip errors 
 and phase errors are both no more than $t$, Alice and 
 Bob will succeed in generating a secure key. In reality, 
 there is no way for two distant observers, Alice and 
 Bob, to verify such a condition directly. Fortunately, 
 Alice and Bob can perform some quality check procedure 
 by randomly sampling their pairs. 
 We have the following Proposition: 
 
 {\bf Proposition~1 (\cite{qkd}, particularly, its supplementary 
 notes VI)}: 
 Suppose Alice prepares $N$ EPR pairs and sends a half of 
 each pair to Bob via a noisy channel (perhaps controlled 
 by Eve). Alice and Bob may randomly select $m$ of those 
 pairs and perform a random measurement along either 
 the $X$ or the $Z$ axis. Suppose, for the moment, 
 that they compute the 
 bit error rates of the tested sample in the two bases 
 separately, thus obtaining $p_X^{sample}$ and $p_Z^{sample}$. 
 Then, these two error rates are good 
 estimates of those of the population (and therefore, 
 also the remaining untested pairs). 
 In particular, one can apply {\it classical} random sampling theory 
 to 
 estimate confidence levels for the error rates in the two bases 
 for the population (and thus the untested pairs). 
 
 {\it Proof of Proposition~1}: 
 Let us summarize the overall strategy of the proof.
One imagines applying the mathematical operation of Bell measurements
on the $N$ imperfect EPR pairs
before the error  correction procedure, but {\it after} Eve's eavesdropping. Consider  the resulting state. It could have been obtained by a different  eavesdropping strategy on the part of Eve, which applies Pauli  operators to the N-EPR-pair state with some probability distribution.  Finally, it suffices to consider only this limited class of eavesdropping  strategies.
 
 Let us consider the state of the
 $N$ EPR pairs after Eve's eavesdropping attack.
 For each of the $m$ tested pair along 
 the 
 $Z$-basis, consider the projection operators, $ P^{i, z}_{||}$ 
 and $ P^{i, z}_{anti-||}$ for the two {\it coarse-grained} outcomes 
 (parallel and anti-parallel) of the measurement performed on the 
 $i$-th pair. 
 Specifically, 
\begin{eqnarray} 
 P^{i,z}_{||} & = & |00\rangle_i \,\langle 00|_i + |11\rangle_i 
\,\langle 11|_i 
\nonumber \\ 
 & = & |\Phi^+\rangle_i \,\langle \Phi^+|_i + |\Phi^-\rangle_i 
\,\langle 
\Phi^-|_i , 
\end{eqnarray} 
\begin{eqnarray} 
 P^{i,z}_{anti-||} & = & |01\rangle_i \,\langle 01|_i + |10\rangle_i 
\,\langle 
 10|_i \nonumber \\ 
 & = & |\Psi^+\rangle_i \,\langle \Psi^+|_i + |\Psi^-\rangle_i 
\,\langle 
\Psi^-|_i , 
\end{eqnarray} 
 where $|\Phi^\pm\rangle = \frac{1}{\sqrt{2}} (|00\rangle \pm 
 |11\rangle)$ and 
 $|\Psi^\pm\rangle = \frac{1}{\sqrt{2}} (|01\rangle \pm |10\rangle)$.

 Similarly, for each of the $m$ test pair along the $X$-axis, consider 
 the projection operators, $ P^{k, x}_{||}$ 
 and $ P^{k, x}_{anti-||}$, for the two {\it coarse-grained} outcomes 
 (parallel and anti-parallel) of the measurement performed on the 
 $k$-th tested pair. Namely, 
\begin{eqnarray} 
 P^{k,x}_{||} & 
 = & \frac{1}{4} ( |0\rangle_k + |1\rangle_k ) \otimes (|0\rangle_k 
 + 
 |1\rangle_k ) 
 (\langle 0|_k + \langle 1|_k) \otimes (\langle 0|_k + \langle 1|_k) \nonumber \\ 
 & & +\frac{1}{4} ( |0\rangle_k - |1\rangle_k ) \otimes (|0\rangle_k
 - 
 |1\rangle_k ) 
 (\langle 0|_k - \langle 1|_k) \otimes (\langle 0|_k - \langle 1|_k) \nonumber \\ 
 & = & |\Phi^+\rangle_k \,\langle\Phi^+|_k 
 + |\Psi^+\rangle_k \,\langle\Psi^+|_k , 
\end{eqnarray} 
\begin{eqnarray} 
 P^{k,x}_{anti-||} & 
 = & \frac{1}{4} ( |0\rangle_k + |1\rangle_k ) \otimes (|0\rangle_k 
 - 
 |1\rangle_k ) 
 (\langle 0|_k + \langle 1|_k) \otimes (\langle 0|_k - \langle 1|_k) \nonumber \\ 
 &  & + \frac{1}{4} ( |0\rangle_k + |1\rangle_k ) \otimes 
 (|0\rangle_k - 
 |1\rangle_k ) 
 (\langle 0|_k + \langle 1|_k) \otimes (\langle 0|_k - \langle 1|_k) \nonumber \\ 
 & = & |\Phi^-\rangle_k \,\langle\Phi^-|_k 
 + |\Psi^-\rangle_k \,\langle\Psi^-|_k . 
\end{eqnarray} 
 
 The above four equations clearly show that 
 using local operations and classical communications only 
 (LOCCs), Alice and Bob can effectively perform a coarse-grained 
 Bell's 
 measurement with these four projection operators. 
 
 Now, consider the operator, $M_B$, which represents a complete 
 measurement 
 along $N$-Bell basis. Since $M_B$, $  P^{i, x}_{||}$, 
 $ P^{i, x}_{anti-||}$, $ P^{k, z}_{||}$ 
 and $ P^{k, z}_{anti-||}$ all refer to a single basis (namely, the 
 $N$-Bell 
 basis), they clearly commute with each other. 
 Therefore, they can be simultaneously diagonalized. Thus, a 
 pre-measurement 
 $M_B$ by say Eve will in no way change the outcome for 
 $ P^{i, x}_{||}$, 
 $ P^{i, x}_{anti-||}$, $ P^{k, z}_{||}$ and $ P^{k, z}_{anti-||}$. 
 Therefore, we may as well consider the case when such a 
 pre-measurement 
 is performed. By doing so, we have reduced the most general eavesdropping  strategy  to a restricted class that involves only Pauli operators.  Consequently, the problem of estimation of the error 
 rates of the two bases is classical. \hfill Q.E.D. 
 
 We emphasize that the key insight of Proposition~1 is the 
 ``commuting observables'' 
 idea: 
 Consider the set of Bell measurements, $X \otimes X$ and 
 $Z \otimes Z$, on all pairs of qubits. 
 All such Bell measurements commute with each other. 
 Therefore, without any loss of generality, 
 we can assign classical probabilities to their 
 simultaneous eigenstates and perform classical 
 statistical analysis. This greatly simplifies the analysis. 
 
 More concretely, provided that total number of the EPR 
 pairs goes to infinity, 
 the classical de Finetti's theorem applies to 
 the random test sample of $m$ pairs. 
 Moreover, for a sufficiently large $N$, it is common in 
 classical statistical theory to assume 
 a normal distribution and use it to estimate 
 the mean of the population and establish confidence levels. 
 Therefore, with a high confidence level, for the remaining 
 untested pairs, the error rates $p_X^{untested} < p_X^{sample} + 
\epsilon$ 
 and $p_Z^{untested} < p_Z^{sample} + \epsilon$. 
 
 The next question is: how do the two error rates (for 
 the $X$ and $Z$ bases) relate to the bit-flip and phase 
 errors in the underlying quantum error correcting code? 
 Suppose, as in our discussion so far, 
 Alice and Bob generate their final 
 key by measuring along the $Z$-axis only. 
 In this case, it should not be hard to see that the 
 bit-flip error has an 
 error rate $p_Z^{untested}$ and the phase error 
 has an error rate $p_X^{untested}$. 
 
 However, in BB84, it is common practice to allow 
 Alice and Bob to generate the key by measuring 
 each pair along either the $X$ or $Z$-axis 
 with uniform probabilities. 
 Mathematically, as discussed in \cite{shorpre,six}, 
 this is equivalent to Alice's applying either i) a Hadamard 
 transform 
 or ii) an identity operator to the qubit before sending it to Bob. 
 Therefore, in this case, it 
 should not be too hard to see that 
 the bit-flip error is given by the averaged 
 error rate $( p_X^{untested} + p_Z^{untested})/2$ of 
 the two bases. Similarly, the phase error rate is 
 given by the same expression. 
 For this reason, 
 it is, in fact, unnecessary in Shor and Preskill's proof 
 for Alice and Bob to compute the two error rates 
 separately. In other words, a simple-minded error 
 analysis in which they lump all polarization data 
 (from both rectilinear or diagonal bases) together and 
 compute a single sample bit error rate, call it $e^{sample}$ 
 is sufficient for 
 the quality check stage. 
 
 Now, suppose a QECC $[[n,k,d]]$ is 
 chosen such that the maximal tolerable error rate, 
 $e^{max} = {t \over n} \equiv {\lfloor { (d - 1)  /2} \rfloor \over 
 n} 
 > e^{sample} + \epsilon$. Then, for any eavesdropping strategy 
 that will pass the quality check stage with a non-negligible 
 probability, it is most likely that the remaining untested 
 $n$ EPR pairs will have less then $t$ bit-flip errors and 
 also less than $t$ phase errors. Therefore, the error 
 correction will most likely succeed and Alice and Bob 
 will share a $k$-EPR-pair state with high fidelity. 
 
 The following theorem shows that once Alice and Bob share 
 a high fidelity $k$-EPR-pair state, then they 
 can generate a key such that the eavesdropper's mutual 
 information is very small. 
 
 {\bf Theorem~2} (\cite{qkd}): Suppose two distant 
 observers, Alice and Bob, share a high fidelity 
 $k$-EPR-pair state, $\rho$, such that 
 $ \langle \Phi^{(k)} | \rho |  \Phi^{(k)}   \rangle > 1 - \delta$ 
 where 
 $\delta \ll 1$ and they generate a key by measuring 
 the state along say the $Z$-axis, then the eavesdropper's mutual 
 information on the key is bounded by 
\begin{equation} 
 S (\rho) < 
 - ( 1 - \delta) \log_2 ( 1 - \delta) - \delta \log_2 { \delta \over  ( 2^{2k}  - 1)} = \delta \times \left( { 1  \over \log_e 2} + 2k + 
\log_2 (  1/ \delta) \right) + O (\delta^2). 
\end{equation}

 {\bf Proof}: Let us recapitulate the proof 
 presented in Section~II of supplementary material of 
\cite{qkd}. The proof consists of two Lemmas. 
 Lemma~A says that high fidelity implies 
 low entropy. Lemma~B says that the entropy is a bound to 
 the eavesdropper's mutual information with Alice and Bob. 
 
 More concretely, Lemma~A says the following: 
 If $ \langle  \Phi^{(k)}| \rho | \Phi^{(k)} \rangle > 
 1 - \delta $ where $\delta \ll 1$, then the von Neumann entropy 
 satisfies $S (\rho) < - ( 1 - \delta) \log_2 ( 1 - \delta) - 
\delta \log_2 { \delta \over ( 2^{2k} -1 )} $. Proof of Lemma~A: 
 If $ \langle \Phi^{(k)}| \rho | \Phi^{(k)} \rangle > 
 1 - \delta $, then the largest eigenvalue of the density matrix 
 $\rho$ must be larger than $ 1 - \delta$. Therefore, 
 the entropy of $\rho$ is, bounded above by that of a 
 density matrix, $\rho_0  = 
 diag ( 1 - \delta, { \delta \over ( 2^{2k} -1 )}, 
 { \delta \over ( 2^{2k} -1 )}, \cdots, { \delta \over ( 2^{2k} -1 )} 
 )$, 
 which has an entropy $- ( 1 - \delta) \log_2 ( 1 - \delta) - 
\delta \log_2 { \delta \over ( 2^{2k} -1 )} $. 
 
 Lemma~B, which is a corollary of Holevo's theorem 
\cite{holevo}, 
 says the following: Given any pure state $\phi_{A'B'}$ of a 
 system consisting of two subsystems, $A'$ and $B'$, and any 
 generalized measurements $X$ and $Y$ on $A'$ and $B'$ respectively, 
 the entropy of each subsystem $S( \rho_{A'})$ (where 
 $\rho_{A'}$ is the reduced density matrix, 
 $Tr_{B'} | \phi_{A'B'} \rangle \langle \phi_{A'B'} | $) is an 
 upper bound to the amount of mutual information between $X'$ and 
 $Y'$. 
 
 Now, Suppose Alice and Bob share a bipartite state 
 $\rho_{AB}$ of fidelity $ 1- \delta$ to $k$ EPR pairs. 
 By applying Lemma~A, 
 one shows that the entropy of $\rho_{AB}$ is 
 bounded by $S (\rho) < - ( 1 - \delta) \log_2 ( 1 - \delta) - 
\delta \log_2 { \delta \over ( 2^{2k} -1 )} $. 
 
 Let us now introduce Eve to the picture 
 and consider the system consisting of the subsystem, $A'$, of Eve 
 and the subsystem, $B'$, of combined Alice-Bob. (i.e., 
 $B' = AB$.) Let us 
 consider the most favorable situation for Eve where she 
 has perfect control over the environment. In this case, 
 the overall (Alice-Bob-Eve) system 
 wavefunction can be described by a pure state, 
 $\phi_{A'B'}$ where Eve controls $A'$ and the combined Alice-Bob 
 controls $B'$. By Lemma~B, Eve's mutual information with 
 Alice-Bob's system is bounded by $( 1 - \delta) \log_2 ( 1 - \delta) 
 - \delta \log_2 { \delta \over ( 2^{2k} -1 )} $. \hfill Q.E.D.

 {\it Remark~1}: It is not too hard to see that Alice and Bob 
 will most likely share a common key that is essentially 
 random in the above procedure. 
 
 {\it Remark~2}: Suppose we limit the eavesdropper's 
 information, $I_{eve}$, to be less than $\epsilon$, 
 Theorem~2 shows that, as the length, $k$ of the 
 final key increases, the allowed infidelity, $\delta$, of 
 the state must decrease at least as $ O (1/k)$.

\subsection{reduction to BB84} 
 
 Shor and Preskill considered a special class of 
 quantum error correcting codes, namely Calderbank-Shor-Steane 
 (CSS) codes. They 
 showed that a QKD that employs 
 an entanglement purification protocol (EPP) based 
 on a CSS code can be reduced to 
 BB84. Let us follow their arguments in two steps. 
 
\subsubsection{from entanglement purification protocol 
 to quantum error-correcting code protocol} 
 
 From the work of BDSW \cite{BDSW}, it is well known that 
 any entanglement purification protocol with only one-way 
 classical communications can be converted into a quantum 
 error-correcting code. Shor and Preskill applied this 
 result to an EPP-based QKD scheme. 
 Let us recapitulate the procedure of an EPP-based QKD scheme. 
 Alice creates $N$ EPR pairs 
 and sends half of each pair to Bob. She then 
 measures the check bits and compares them with Bob. 
 If the error rate is not too high, Alice then measures 
 $M_{i,A}$ and publicly announces the outcomes to Bob, who 
 measures $M_{i,B}$. 
 This allows Alice and Bob to correct errors and distill out 
 $k$ perfect EPR pairs. Alice and Bob then measure 
 $\bar{Z}_{a,A}$ and $\bar{Z}_{a,B}$, the encoded $Z$ operators, to 
 generate the key. 
 
 Note that, by locality, it does not matter whether Alice measures the 
 check bits before or after she transmits halves of EPR pairs to 
 Bob. Similarly, it does not matter whether Alice measures 
 her syndrome (i.e., the stabilizer elements, $M_{i,A}$) before or 
 after the 
 transmission. Now, if she measures her check bits before 
 the transmission, it is equivalent to choosing a random 
 BB84 state, 
 $ | 0 \rangle, | 1 \rangle, | + \rangle = { 1 \over \sqrt{2}} 
 ( | 0 \rangle + | 1 \rangle ), | - \rangle = 
 { 1 \over \sqrt{2} 
 } ( | 0 \rangle - | 1 \rangle )$. 
 If Alice measures her syndromes before the transmission, it is 
 equivalent to encoding halves of $k$ EPR pairs in an $[[n,k,d]]$ 
 QECC, ${\cal C}_{s_A}$, and sending them to Bob, 
 where ${\cal C}_{s_A}$ is the corresponding quantum code for the 
 syndrome, $s_A$, she found. 
 
 Finally, suppose Alice measures her halves of the encoded $k$ EPR 
 pairs 
 before the transmission, it is equivalent to 
 Alice preparing one of the $2^k$ mutually orthogonal codeword states 
 in the quantum code, ${\cal C}_{s_A}$, to 
 represent a $k$-bit key and sending the state to Bob. 
 In summary, the above discussion reduces a QKD protocol based on 
 EPP to a QKD protocol based on a class of $[[n,k,d]]$ 
 QECC, ${\cal C}_{s_A}$'s. 
 
\subsubsection{from error-correcting protocol to BB84} 
 So far, we have not specified which class of QECCs to employ. 
 Notice that, for a general QECC, the QECC protocol still requires 
 quantum computers to implement (for example, the operators 
 $M_{i,A}$). 
 Here comes a key insight of Shor and Preskill: If one 
 employs Calderbank-Shor-Steane 
 (CSS) codes \cite{CS,steane}, then the scheme can be further reduced 
 to 
 standard BB84, which can be implemented {\it without} 
 a quantum computer. 
 CSS codes have the nice property that the bit-flip and 
 phase error correction procedures are totally decoupled 
 from each other. In other words, the error syndrome is 
 of the form of a pair $(s_b, s_p)$ where, $s_b$ and $s_p$ are 
 respectively the bit-flip and phase error syndrome. 
 Without quantum computers, there is no way for Alice and 
 Bob to compute the phase error syndrome, $s_p$. However, this is 
 not really a problem because phase errors do not change 
 the value of the final key, which is all that Alice and Bob 
 are interested in. For this reason, Alice and Bob can 
 basically drop the phase-error correction procedure. 
 
 Let us first introduce the CSS code. 
 Consider two classical binary codes, 
 $C_1$ and $C_2$, such that, 
\begin{equation} 
\{ 0 \} \subset C_2 \subset C_1 \subset F^n_2, 
\label{e:CSScodes1} 
\end{equation} 
 where $F^n_2$ is the binary vector space of the $n$ bits 
 and that both $C_1$ and $C_2^{\perp}$, the dual of $C_2$, 
 have a minimal distance, $d= 2t +1$, for some integer, $t$. 
 The basis vectors of a CSS code, $\cal C$, are: 
\begin{equation} 
 v \to | \psi(v) \rangle= 
 { 1 \over | C_2|^{1/2}} \sum_{w \in C_2} | v + w \rangle, 
\label{e:CSScodes2} 
\end{equation} 
 where $v \in C_1$. Note that, whenever $v_1 - v_2 \in C_2$, 
 they are mapped to the same state. In fact, the basis vectors 
 are in one-one correspondence with the cosets of $C_2$ in $C_1$. 
 The dimension of a CSS code is $2^k$ where $k = {dim (C_1) - dim 
 (C_2)}$. 
 In standard QECC convention, the CSS code 
 is denoted as an $[[n,k,d]]$ QECC. 
 
 One can also construct a whole class of CSS codes, ${\cal C}_{z,x}$, 
 from $\cal C$, where the basis vectors of ${\cal C}_{z,x}$ are of 
 the form 
\begin{equation} 
 v \to  | \psi(v)_{z,x} \rangle= 
 { 1 \over | C_2|^{1/2}} \sum_{w \in C_2} ( -1)^{x \cdot w} 
 | v + w + z \rangle, 
\label{e:cssshifted} 
\end{equation} 
 where $v \in C_1$.\footnote{Note 
 that our notation is different from both Refs. \cite{shorpre} 
 and \cite{squeeze} in that we have interchanged 
 $x$ and $z$ in Eq.~(\ref{e:cssshifted}) as well as in 
 the definition of ${\cal C}_{z,x}$. 
 In our notation, $z$ denotes the bit-flip error syndrome 
 and $x$ denotes the phase error syndrome.} 
 
 Let us introduce some notation. Recall the definition of 
 Pauli matrices. The operator 
 $\sigma_x$ corresponds to 
 a bit-flip error, $\sigma_z$ a phase error and $\sigma_y$ a 
 combination of 
 both bit-flip and phase errors. 
 It is convenient to denote the Pauli operator acting on the $k$-th 
 qubit 
 by $\sigma_{a(k)}$ where, $a \in \{x, y, z\}$. 
 Given a binary vector $s \in F^n_2$, let 
\begin{equation} 
\sigma^{[s]}_a 
 = \sigma^{s_1}_{a(1)} \otimes \sigma^{s_2}_{a(2)} \otimes \cdots 
\sigma^{s_n}_{a(n)} . 
\end{equation} 
 By definition, the eigenvalues of $\sigma^{[s]}_a$ are $+1$ and 
 $-1$. 
 
 Let $H_1$ be the parity check matrix for the code $C_1$ and 
 $H_2$ be the parity check matrix for $C_2^{\perp}$. 
 For each row, $r \in H_1$, consider an operator, 
 $\sigma^{[r]}_z$. Applying to a quantum state, 
 their simultaneous eigenvalues give the bit-flip error syndrome. 
 For each row, $s \in H_2$, consider an operator, 
 $\sigma^{[s]}_x$. Applying to a quantum state, 
 their simultaneous eigenvalues give the phase error syndrome. 
 For instance, when applied to the state, 
 $\psi (v)$ in Eq.~(\ref{e:cssshifted}), 
 we find the bit-flip error syndrome, $s_b$, and the phase error 
 syndrome, $s_p$ to be: 
\begin{equation} 
\begin{array}{cc} 
 s_b = H_1 (z), 
 & 
 s_p = H_2 (x). 
\end{array} 
\end{equation} 
 
 Let us look at the QECC-based QKD scheme as a whole. 
 Alice is supposed to 
 pick a random vector $v \in C_1$, random $x_A$ and $z_A$ and encode 
 it 
 as  $| \psi(v)_{z_A,x_A} \rangle$. 
 After Bob's acknowledgement of his receipt of 
 the state, Alice 
 then announces the values of $x_A$ and $z_A$ to Bob. Bob measures 
 the state 
 and obtains his own syndrome, the values of $x_B$ and $z_B$. The 
 relative syndrome, the values of $x_A \times x_B$ and $z_A \times 
 z_B$, is 
 the actual error syndrome of the channel. Bob then corrects 
 the errors and measures along the $z$-axis to obtain a string $v + w 
 + z_A$ 
 for some $w \in C_2$. 
 He then subtracts $x_A$ to obtain $v+ w$. 
 Finally, Bob applies the generator matrix\footnote{Gottesman 
 and Preskill's paper 
 stated that the parity check matrix, $H_2$, of 
 the dual code $C^{\perp}_2$ should be used. But, it should really 
 be the generator matrix.}, $G_2$, of the 
 dual code $C^{\perp}_2$ (i.e., 
 the parity check matrix of the 
 code $C_2$) to generate the key, 
\begin{equation} 
 G_2  (v + w) = G_2 (v) + G_2 (w) = G_2 (v). 
\label{e:keygeneration} 
\end{equation} 
 Notice that the key is in one-one correspondence with the coset 
 $C_2$ in 
 $C_1$ because of the mapping $G_2 (v) \to v + C_2$.\footnote{This is  a well-known result in classical coding theory.} 
 
 Here comes the key point: 
 Since Bob measures along the $z$-axis to generate the key, the phase 
 errors really do not change the value of the key. Therefore, 
 it is not necessary for Alice to announce the phase error syndrome, 
 $x_A$, to Bob. Therefore, without affecting the security of the 
 scheme, Alice is allowed to prepare a state $\psi (v)_{z_A,x_A}$ 
 and then discard, rather than broadcast the value of $x_A$. 
 Equivalently, she is allowed to prepare an {\it averaged} state 
 $\psi (v)_{z_A, x_A}$ over all values of $x_A$. The 
 averaging operation destroys the phase coherence and, from 
 Eq.~(\ref{e:cssshifted}), leads to 
 a classical mixture of $| v + w + z_A \rangle$ in the $z$-basis. 
 
 As a whole, the error correction/privacy amplification procedure for 
 the resulting BB84 QKD scheme goes as follows: 
 Alice sends $ | u \rangle$ to Bob through a quantum channel. 
 Bob obtains $ u + e$ due to channel errors. Alice 
 later broadcasts $u + v$, for a random $v \in C_1$. 
 Bob subtracts it from his 
 received string to obtain $ v + e$. He corrects the errors 
 using the code $C_1$ to obtain a codeword, $v \in C_1$. 
 He then applies the matrix, $G_2$, to generate the final key 
 $G_2 (v)$, which is in one-one correspondence with a coset of 
 $C_2$ in $C_1$. 
 
 {\it Remark~3}: Upon reduction from CSS code to 
 BB84, the original bit-flip error correction procedure 
 of $C_1$ becomes 
 a classical error correction procedure. On the other hand, 
 the phase error correction procedure becomes a privacy amplification 
 procedure. (And, it is achieved by extracting the coset of $C_2$ in 
 $C_1$ 
 by using the generator matrix, $G_2$, of the dual code 
 $C^{\perp}_2$.) 
 
 {\it Remark~4}: Note that the crux of this reduction is to 
 demonstrate that Eve's view 
 in the original EPP picture can be made to be exactly 
 the same as in BB84. 
 Therefore, the fact that Alice and Bob {\it could have} executed 
 their QKD with quantum computers is sufficient to guarantee 
 the security of QKD. They do not actually need quantum computers 
 in the actual execution. Another way to saying what is going on is 
 that Alice and Bob are allowed to {\it throw away} the phase error 
 syndrome information without weakening security. By throwing 
 such phase error syndrome away, the scheme becomes implementable 
 with only classical computers, and, therefore, 
 does not require quantum computers. 
 
\subsection{Acceptable error rate} 
 If one only aims to 
 decode noise patterns up to half of the minimal distance $d$ 
 (as in much of conventional coding theory), then, given 
 that above quantum code uses 
 $C_1$ and $C^{\perp}_2$ that have large minimal 
 distances, it achieves the quantum Gilbert-Varshamov bound 
 for CSS codes\cite{CS,steane}. As the length of the code, 
 $n$ goes to infinity, the number of encoded qubits goes to 
 $ [ 1 - 2H (2 e ) ]n $, where $e$ is the measured bit error rate
in the quantum transmission. Here, the factor of $2$ in front of 
 $H$ arises because one has to deal with both phase and bit-flip 
 errors in a quantum code. In the classical analog, the 
 factor of $2$ in front of $H$ does not appear. 
 (The factor of $2$ inside $H$ ensures 
 that the distance between any two codewords is at least twice of 
 the tolerable error rate.) 
 
 However, in fact, the same 
 CSS code can decode, with vanishing probability of error, 
 up to {\it twice} of the above error rate. That is to say, it 
 can achieve the quantum Shannon bound for 
 {\it non-degenerate} codes. Asymptotically, 
 the number of encoded qubits goes to 
 $ [ 1 - 2 H ( e ) ] n$. The maximal tolerable 
 error rate would be about $11 \%$. 
 
 The reason for the improvement is 
 that the code only needs to correct the {\it likely} errors, 
 rather than all possible errors at such a noise level. 
 We remark that this is highly reminiscent of a result 
 in classical coding theory which states that Gallager 
 codes, which are based 
 on very low density parity check matrices, can achieve the 
 Shannon bound in classical coding theory\cite{MacKay}. 
 In the classical case, the intuition 
 is that in a very high-dimensional binary 
 space, while two spheres of radius $r$ whose centers are 
 a distance $d$ apart have a non-zero volume of intersection 
 for any $r$ greater than $d/2$, the 
 {\it fractional} overlap is {\it vanishingly small} provided that 
 $r < d$. 
 
 To achieve the Shannon bound in the quantum code case, 
 it is necessary to ensure that the errors are randomly 
 distributed among the $n$ qubits. As noted by 
 Shor and Preskill, this can be done 
 by, for example, permuting the $n$ qubits randomly. 
 
 {\it Remark~5}: In 
 the original Mayers' proof, the maximal tolerable error rate is 
 about $7\%$. As noted by Shor and Preskill, Mayers' proof has 
 a hidden CSS code structure. Mayers considered some 
 (efficiently decodable) classical 
 codes, $C_1$, and a random subcode, $C_2$, of $C_1$. 
 It turns out that, the dual, $C^{\perp}_2$, of a random subcode of 
 $C_1$ is highly likely to be a good code. 
 However, Mayers' proof considered the correction of {\it all} 
 phase errors, rather than {\it likely} phase errors within 
 the error rate. For this reason, as the length, $n$, of the 
 codeword goes to infinity, the number of encoded qubits 
 asymptotically 
 approaches 
 $ [ 1 - H( e ) - H ( 2 e ) ] n$, 
 the first $H$ comes from error correction and the 
 second comes from privacy amplification. Thus, key generation is 
 possible only up to $7\%$. Shor and Preskill extended 
 Mayers' proof by noting that it is necessary to 
 correct only {\it likely} phase errors, but not all phase errors 
 within the error rate. They also randomize the errors 
 by adding the permutation step 
 mentioned in the above paragraphs. 
 
\subsection{Shor and Preskill's protocol of BB84} 
\label{subsec:shorpreskill} 
 In the last few subsections, we have already discussed 
 the main steps of Shor and Preskill's proof. 
 For completeness, we will list here all the steps of 
 Shor and Preskill's protocol of BB84 scheme. 
 
 (1) Alice sends a sequence of say $(4 + \delta_1 )n $, where
$\delta_1$ is a small positive number,
 photons each in one of the 
 four polarizations (horizontal, vertical, 45 degrees and 135 
 degrees) 
 chosen randomly and independently. 
 
 (2) For each photon, Bob chooses the type 
 of measurement randomly: along either the rectilinear or diagonal 
 bases. 
 
 (3) Bob records his measurement bases and the results of the 
 measurements. 
 
 (4) Subsequently, 
 Bob announces his bases (but {\em not\/} the results) through 
 the public unjammable channel that he shares with Alice. 
 
 {\it Remark~6}: Notice that it is crucial that Bob announces his 
 basis only after 
 his measurement. This ensures that during the 
 transmission of the signals through the quantum channel the 
 eavesdropper Eve 
 does not know which basis to eavesdrop along. 
 Otherwise, Eve can avoid detection simply by 
 measuring along the same basis used by Bob. 
 
 (5) Alice tells Bob which of his measurements have been done in the 
 correct 
 bases. 
 
 (6) Alice and Bob divide up their polarization data into two classes 
 depending on whether they have used the same basis or not. 
 
 {\it Remark~7}: 
 Notice that on average,
 Bob should have performed the wrong type of measurements on  half of the photons. Here, by a wrong type of measurement 
 we mean that Bob has used a basis different from that of Alice. 
 For those photons, he gets random outcomes. 
 Therefore, he throws away those polarization data. 
 We emphasize that this immediately implies that half of the 
 data are thrown away and the efficiency of BB84 is bounded by 50\%. 
 
 With high probability, at least $\approx 2n$ photons are left. (If 
 not, 
 they abort.) Assuming that no eavesdropping has occurred, 
 all the photons 
 that are measured by Bob in the correct bases should give the 
 same polarizations as prepared by Alice. Besides, 
 Bob can determine those polarizations by his own detectors without 
 any communications from Alice. Therefore, 
 those polarization data are a candidate for their raw key. 
 However, before they proceed any further, it is 
 crucial that they test for tampering. 
 For instance, they can use the following 
 simplified method for estimating the error rate. (Going 
 through BB84 would give us essentially the same result, namely that 
 all 
 accepted data 
 are lumped together to compute a 
 {\em single\/} error rate.) 
 
 (7) Alice and Bob randomly pick a subset of photons from those 
 that are measured in the correct bases and publicly compare 
 their polarization data for preparation and measurement. 
 For instance, they can use $\approx n$ photons for such testing. 
 For those results, they estimate the error rate for the 
 transmission. 
 Of course, since the polarization data of photons 
 in this subset have been announced, Alice and Bob must 
 sacrifice those data to avoid information leakage to Eve. 
 
 We assume that Alice and Bob have some idea on the channel 
 characteristics. 
 If the average error rate $\bar{e}$ turns out to be unreasonably 
 large 
 (i.e., $\bar{e} \geq e_{\rm max}$ where 
 $e_{\rm max}$ is the maximal tolerable error rate), then either 
 substantial eavesdropping has occurred or the channel is somehow 
 unusually noisy. In both cases, all the data are discarded 
 and Alice and Bob may re-start the whole procedure again. 
 Notice that, even then there is no loss in security 
 because the compromised key is never used to encipher sensitive 
 data. 
 Indeed, Alice and Bob will derive a key from the data only when 
 the security of the polarization data is first established. 
 
 On the other hand, if the error rate turns out to be reasonably 
 small 
 (i.e., $\bar{e} < e_{\rm max}$), they go to 
 the next step. 
 
 (8) Reconciliation and privacy amplification: Alice and Bob can 
 independently convert the polarizations of the remaining $n$ 
 photons into a {\em raw\/} key by, for example, 
 regarding a horizontal or 45-degree photon as 
 denoting a `0' and a vertical or 135-degree photon a `1'. 
 
 Alice and Bob pick a CSS code based on 
 two classical binary codes, $C_1$ and $C_2$, 
 as in Eqs.~(\ref{e:CSScodes1}) and (\ref{e:CSScodes2}), 
 such that both $C_1$ and $C_2^{\perp}$, the dual of $C_2$, correct 
 up to $t$ errors where $t$ is chosen such that 
 the following procedure of error correction and 
 privacy amplification will succeed with a high probability.

 (8.1) Let $v$ be Alice's string of the remaining $n$ unchecked bits. 
 
 Alice picks a random codeword $u \in C_1$ and publicly announces 
 $u + v$. 
 
 (8.2) Let $v + \Delta $ be Bob's string of the 
 remaining $n$ unchecked bits. (It differs from Alice's string 
 due to the presence of errors $\Delta $.) Bob subtracts Alice's 
 announced string $u + v$ from his own string to obtain $u + \Delta$, 
 which is a corrupted version of $u$. Using the error correcting 
 property of $C_1$, Bob recovers a codeword, $u$, in $C_1$. 
 
 (8.3) Alice and Bob use the coset of $u + C_2$ as their key. 
 
 {\it Remark~8}: As noted before, there is a minor subtlety 
\cite{shorpre}. 
 To tolerate a higher channel error rate of up to about $11\%$, 
 Alice should apply a random permutation to the qubits before 
 their transmission to Bob. Bob should then apply the inverse 
 permutation before decoding. 
 
 {\it Remark~9}: Depending on the desired security level, 
 the number of test photons in Step (7) can be made to be much 
 smaller than 
 $n$. If one takes the limit that the probability that Eve can 
 break the system is fixed but arbitrary, 
 then the number of test photons can be 
 made to be of order $\log n$ only. On the other hand, if 
 the probability that Eve can break the system is chosen to be 
 exponentially small in $n$, then it is necessary to test 
 order $n$ photons.

\section{Overview of efficient BB84} 
\label{sec:overview} 
 In this section, we will give 
 an overview of the efficient BB84 scheme 
 and provide a sketch of a simple proof of its security. 
 
\subsection{bias}
 The first major new ingredient of 
 our efficient BB84 scheme is to put a bias in the probabilities of 
 choosing between the two bases. 
 
 Recall the fraction of rejected data of BB84 is likely to be at 
 least $50\%$. 
 This is because in BB84 
 Alice and Bob choose between the two bases randomly and 
 independently. 
 Consequently, on average Bob performs a wrong type of measurement 
 half of the time and, therefore, half of the photons are thrown away 
 immediately. The efficiency will be increased if Alice prepares and  Bob measures their photons with a biased choice of basis.  Specifically, they first agree on a fixed number $0< p \leq 1/2$. Alice 
 prepares (Bob measures) each photon randomly, independently in the rectilinear  and diagonal basis with probabilities $p$ and $1-p$ respectively.  Clearly, the scheme is insecure when $p=0$. Nonetheless, we shall show that  in the limit of large number of photon transfer, this biased scheme is  secure in the limit of $p\rightarrow 0^+$. Hence, the efficiency of this  biased scheme is asymptotically doubled when compared to BB84.

 Notice also that the bias in the probabilities might be produced 
 passively 
 by an apparatus, for example, an unbalanced beamsplitter in 
 Bob's side. Such a passive 
 implementation based on a beamsplitter 
 eliminates the need for fast switching 
 between different polarization bases and is, thus, useful in 
 experiments. 
 This may not be obvious to the readers why a beamsplitter can 
 create a probabilistic implementation. If one uses 
 a beamsplitter, rather than a fast switch, one gets a 
 superposition of states and not a mixture. However, provided that 
 the subsequent measurement operators annihilate any 
 state transmitting in one of the two paths, the probabilities of 
 the outcomes will be the same for either a mixture or a 
 superposition. 
 More concretely, suppose one can model the 
 problem by decomposing the Hilbert space into two 
 subspaces ${\cal H} = {\cal H}_1 \oplus {\cal H}_2 $ where 
 ${\cal H}_1$ is the Hilbert subspace corresponding to the 
 first path and ${\cal H}_2$ the second respectively. 
 Consider the two sets of measurement operators, $\{P_i\}$'s 
 and $\{Q_j \}$'s respectively, where $P_i | \psi \rangle = 0$ 
 for all $  | \psi \rangle \in {\cal H}_2$ and 
 $Q_j | \psi \rangle = 0$ 
 for all $  | \psi \rangle \in {\cal H}_1$. Let us write 
 $ | u \rangle =  | u_1 \rangle +  | u_2 \rangle $ where 
 $ | u_1 \rangle \in {\cal H}_1$ and $ | u_2 \rangle \in {\cal H}_2$. 
 
 Now, the probability of the outcome corresponding to the measurement 
$P_i$ is given by 
\begin{equation} 
 | \langle u | P_i |  u \rangle| = 
 | \langle u_1 | P_i |  u_1\rangle |
\end{equation} 
 and the 
 probability of the outcome corresponding to the measurement 
 $Q_j$ is given by 
\begin{equation} 
 | \langle u | Q_j |  u \rangle | = 
 | \langle u_2 | Q_j |  u_2 \rangle | . 
\end{equation} 
 Those probabilities are exactly the same as those given by 
 a mixture of $ |u_1 \rangle$ and $| u_2 \rangle$. 
 
\subsection{Refined Error Analysis} 
 In the original BB84 scheme, all the accepted data (those 
 for which Alice and Bob measure along the same basis) are lumped 
 together to compute a {\em single\/} error rate. In this subsection, 
 we introduce the second major ingredient of our scheme --- a 
 refined error analysis. The idea is for Alice and Bob to 
 divide up the accepted data into two subsets according to 
 the actual basis (rectilinear or diagonal) used. After that, 
 a random subset of photons is drawn from each of the two sets. 
 They then publicly compare their polarization data and 
 from there estimate the error rate for each basis {\em separately}. 
 They decide that the run is acceptable if and only if both error 
 rates are sufficiently small. 
  
 The requirement of having estimated error rates separately in
 both bases to be small is more stringent that the original one.
 In fact, if a naive 
 data analysis, where only a single error rate is 
 computed by Alice and Bob, had been employed, our new scheme 
 would have been insecure. 
 To understand this point, consider the following example 
 of a so-called biased 
 eavesdropping strategy by Eve. 
 
 For each photon, Eve 1) with a probability $p_1$ 
 measures its polarization along the rectilinear basis 
 and resends the result of her measurement to Bob; 
 2) with a probability $p_2$ measures its polarization along the 
 diagonal basis and resends the result of her measurement to Bob; and 
 3) 
 with a probability $1-p_1 -p_2$, does nothing. 
 We remark that, by varying the values of $p_1$ and $p_2$, 
 Eve has a whole class of eavesdropping strategies. 
 Let us call any of the strategies in this class 
 a biased eavesdropping attack. 
 
 Consider the error rate $e_1$ for the case when both Alice and 
 Bob use the rectilinear basis. For the biased eavesdropping strategy 
 under current consideration, 
 errors occur only if Eve uses the diagonal basis. 
 This happens with a {\em conditional\/} probability $p_2$. 
 In this case, 
 the polarization of the photon is randomized, thus giving an 
 error rate $e_1 = p_2/2$. 
 Similarly, errors for the diagonal basis 
 occur only if Eve is measuring along the 
 rectilinear basis. This happens with a conditional 
 probability $p_1$ and when it happens, the photon polarization is 
 randomized. Hence, 
 the error rate for the diagonal basis $e_2 = p_1/2$. 
 Therefore, Alice and Bob 
 will find, for the biased eavesdropping attack, 
 that the average error rate 
\begin{equation} 
\bar{e} = { p^2 e_1 + ( 1 - p)^2 e_2  \over 
 p^2 +  ( 1 - p)^2 } 
 = { p^2 p_2 + ( 1 -p )^2 p_1  \over 
 2 [ p^2 +  ( 1 - p)^2] } . 
\end{equation} 
 
 Suppose Eve always eavesdrops solely along the diagonal basis (i.e., 
 $p_1 =0$ and $p_2 = 1$), then 
\begin{equation} 
\bar{e} = { p^2   \over 
 2 [ p^2 +  ( 1 - p)^2] }  \to 0 
\end{equation} 
 as $p$ tends to $0$. 
 Hence, with the original error estimation method 
 in BB84, Alice and Bob will fail to detect eavesdropping by Eve. 
 Yet, Eve will have much information about Alice and Bob's raw key 
 as she is always eavesdropping along the dominant (diagonal) basis. 
 Hence, a naive error analysis fails miserably. 
 
 In contrast, the refined error analysis
 can make our scheme secure against such a biased eavesdropping 
 attack. Recall that in a refined error analysis, the two error rates 
 are computed {\em separately}. The key observation is that 
 these two error rates $e_1= p_2/2$ and $e_2= p_1/2$ depend 
 only on Eve's eavesdropping strategy, but {\em not\/} on 
 the value of $\epsilon$. This is so because 
 they are {\em conditional\/} probabilities. 
 Consequently, in the case that Eve is always eavesdropping along
 the dominant (i.e., diagonal) basis, Alice and Bob will find an error rate of 
 $e_1= p_2/2 = 1/2$ for the rectilinear basis. 
 Since $1/2$ is substantially larger than $e_{max}$, Alice 
 and Bob will successfully catch 
 Eve. 

\subsection{Procedure of efficient QKD} 
 We now give the complete procedure of an efficient QKD scheme. 
 Its security will be discussed in Subsection~\ref{ss:proofeff} 
 and more details of a proof of its security will be given 
 in Section~\ref{s:proof}. 
 
 {\bf Protocol~E: Protocol for efficient QKD} 
 
 (1) Alice and 
 Bob pick a number $0 < p \leq 1/2$ whose value is 
 made public. Let N be a large integer.
Alice sends a sequence of N photons to Bob. 
 For each photon Alice chooses between 
 the two bases, rectilinear and diagonal, with probabilities 
 $p$ and $1 -p $ respectively. 
 The value of $p$ is chosen 
 so that $N (p^2 - \delta') = m_1 = \Omega ( \log N) $, 
 where $\delta'$ is some small positive number and 
 $m_1$ is the number of test photons in the rectilinear basis 
 in Step~(7). 
 
 (2) Bob measures the polarization of each received photon 
 independently along the rectilinear and diagonal bases 
 with probabilities $p$ and $1-p$ respectively. 
 
 (3) Bob records his measurement bases and the results of the 
 measurements. 
 
 (4) Bob announces his bases (but {\em not\/} the results) through 
 the public unjammable channel that he shares with Alice.

 (5) Alice tells Bob which of his measurements have been done in the 
 correct 
 bases. 
 
 (6) Recall that each of Alice and Bob 
 uses one of the two bases --- rectilinear and diagonal. 
 Alice and Bob divide up their 
 polarization data into four cases according to the actual bases 
 used. 
 They then throw away the two cases 
 when they have used different bases. The remaining two cases 
 are kept for further analysis. 
 
 (7) From the subset where they both use the rectilinear basis, Alice 
 and Bob randomly pick a fixed number say $m_1$ photons and publicly 
 compare 
 their polarizations. (Since $N (p^2 - \delta') = m_1$, for a large 
 $N$, 
 it is highly likely that at least $m_1$ photons are transmitted 
 and received in the rectilinear basis. If not, they abort.) 
 The number of mismatches $r_1$ tells them 
 the estimated error rate $e_1 = r_1/m_1$. 
 Similarly, from the subset where they both use the diagonal basis, 
 Alice and Bob randomly pick a fixed number 
 say $m_2$ photons and publicly compare 
 their polarizations. The number of mismatches $r_2$ gives 
 the estimated error rate $e_2 = r_2/ m_2$. 
 
 Provided that the test samples $m_1$ and $m_2$ are sufficiently 
 large, 
 the estimated error rates $e_1$ and $e_2$ should be rather 
 accurate. 
 As will be given in Subsection~\ref{ss:summary}, $m_1$ and 
 $m_2$ should be at least of order $\Omega ( \log k)$, where
$k$ is the length of the final key.
 Now they demand that $e_1, e_2 < e_{\rm max} - \delta_e $ where 
 $e_{\rm max}$ 
 is a prescribed maximal tolerable error rate and $\delta_e$ is some 
 small positive parameter. 
 If these two independent constraints are satisfied, they proceed to 
 step (8). Otherwise, they throw away the polarization data and 
 re-start the whole procedure from step (1).

 (8) Reconciliation and privacy amplification: 
 For simplicity, in what follows, we will take $m_1 =m_2 =N (p^2 - 
\delta')  $. 
 Alice and Bob randomly pick $n= N [ ( 1- p)^2 - p^2 -\delta' ]$ 
 photons from those untested photons 
 that are transmitted and received in the diagonal 
 basis. Alice and Bob then 
 independently convert the polarizations of those $n$ 
 photons into a {\em raw\/} key by, for example, 
 regarding a 45-degree photon as 
 denoting a `0' and a 135-degree photon a `1'. 
 
 {\it Remark~10}: Note that the raw key is generated by measuring 
 along a single basis, namely the diagonal basis. 
 This greatly simplifies the analysis without compromising 
 efficiency or security. 
 
 Alice and Bob pick a CSS code based on 
 two classical binary codes, $C_1$ and $C_2$, 
 as in Eqs.~(\ref{e:CSScodes1}) and (\ref{e:CSScodes2}), 
 such that both $C_1$ and $C_2^{\perp}$, the dual of $C_2$, correct 
 up to $t$ errors where $t$ is chosen such that 
 the following procedure of error correction and 
 privacy amplification will succeed with a high probability. 
 
 (8.1) Let $v$ be Alice's string of the remaining $n$ unchecked bits. 
 
 Alice picks a random codeword $u \in C_1$ and publicly announces 
 $u + v$. 
 
 (8.2) Let $v + \Delta $ be Bob's string of the 
 remaining $n$ unchecked bits. (It differs from Alice's string 
 due to the presence of errors $\Delta $.) Bob subtracts Alice's 
 announced string $u + v$ from his own string to obtain $u + \Delta$, 
 which is a corrupted version of $u$. Using the error correcting 
 property of $C_1$, Bob recovers a codeword, $u$, in $C_1$. 
 
 (8.3) Alice and Bob use the coset of $u + C_2$ as their key. 
 
 {\it Remark~11}: As noted before, there is a minor subtlety 
\cite{shorpre}. 
 To tolerate a higher channel error rate of up to about $11\%$, 
 Alice should apply a random permutation to the qubits before 
 their transmission to Bob. Bob should then apply the inverse 
 permutation before decoding.

\subsection{Outline proof of Security of efficient QKD scheme} 
\label{ss:proofeff} 
 In this subsection, we will give the general strategy of proving 
 the unconditional security of 
 efficient QKD scheme and discuss some subtleties. 
 Some loose ends will be tightened in Section~\ref{s:proof}. 
 
 First of all, we would like to derive the relationship 
 between the error rates in the two bases ($X$ and $Z$) in biased 
 BB84 and 
 the bit-flip and phase error rates in the underlying entanglement 
 purification protocol (EPP). Actually, this depends on 
 how the key is generated. If the key is generated only 
 from polarization data in say the $Z$-basis, then 
 clearly, the bit-flip error rate is simply the $Z$-basis 
 bit error rate and the phase error rate is simply the 
 $X$-basis bit error rate. 
 On the other hand, if the key is generated only from polarization 
 data in say the $X$-basis, then the bit-flip error rate is 
 simply the $X$-basis 
 bit error rate and the phase error rate is simply the 
 $Z$-basis bit error rate. 
 
 More generally, if a key is generated by making a fraction, $q$, 
 of the measurements along the $Z$-basis and a fraction, $1-q$, 
 along the $X$-basis, then the bit-flip and phase error 
 rates are given by weighted averages of the bit error 
 rates of the two bases: 
\begin{eqnarray} 
 e^{bit-flip} &=&  q e_1 + ( 1- q) e_2  \nonumber \\ \cr 
 e^{phase} &=& q  e_1 + ( 1-q)e_2 , 
\label{eq:bitflip} 
\end{eqnarray} 
 where $e_1$ and $e_2$ are the bit error rates of the 
 $Z$ and the $X$ bases respectively. 
 
 Now, in a refined data analysis, Alice and Bob separate data from 
 the two bases into two sets and compute the error rates in the 
 two sets 
 individually. This gives them individual estimates on 
 the bit error rates, $e_1$ and $e_2$, of the $Z$ and $X$ bases 
 respectively. They 
 demand that both error rates must be sufficiently small, say, 
\begin{equation} 
 0 \leq e_1 , e_2 < e_{max}  - \delta_e . 
\label{eq:channel} 
\end{equation} 
 
 From Eqs. (\ref{eq:bitflip}), we see that, 
 provided that the bit error rates of the 
 $X$ and $Z$ bases are sufficiently small 
 (such that Eqs.\ (\ref{eq:channel}) are satisfied), 
 we have 
\begin{equation} 
 0 \leq e^{bit-flip}, e^{phase}  < 11\% , 
\label{eq:signal} 
\end{equation} 
 which says that 
 both bit-flip and phase-flip signal error rates of 
 the underlying EPP are small enough to 
 allow CSS code to correct. Therefore, Shor and Preskill's argument 
 carries over directly to establish the security of 
 our efficient QKD scheme, if Alice and Bob apply a refined 
 data analysis. This completes our sketch of the proof of security. 
 
 We remark that the error correction and privacy amplification 
 procedure that we use are exactly the same as in Shor-Preskill's 
 proof. The point is the following: Once the error rate for 
 both the bit-flip and phase errors are shown to be correctable 
 by a quantum (CSS) code, the procedure for error correction and 
 privacy 
 amplification in their proof can be carried over directly to 
 our new scheme. 
 
\subsection{practical issues} 
\label{ss:practical} 
 Several complications deserve attention. 
 First, Alice and Bob 
 only have estimators of $e_1$ and $e_2$, the bit error 
 rates of the two bases, from their random sample. 
 They need to establish confidence levels on the actual bit 
 error rates of the population (or more precisely, 
 those of the {\it untested} signals) from those 
 estimators.   
 Second, Alice and Bob are interested in 
 the bit-flip and phase error rates of the EPP, rather than 
 the bit error rates of the two-bases. Some 
 conversion of the confidence levels has to be done. 
 Given that the two bases are weighted differently, such a 
 conversion looks non-trivial. 
 Third, Alice and Bob have to deal with finite 
 sample and population sizes whereas many 
 statistics textbooks takes the limit of infinite 
 population size. Indeed, it is commonplace in 
 statistics textbooks to take the limit of infinite 
 population size and, therefore, assume a normal distribution. 
 Furthermore, in practice, Alice and Bob 
 are interested in bounds, not approximations 
 (which might over-estimate or under-estimate) 
 which many statistics textbooks are contented with. 
 
 Another issue: it is useful to specify the constraints on the bias 
 parameter, $q$, and the size of the test samples, $m_1$ and $m_2$. 
 Indeed, in order to demonstrate the security of 
 an efficient scheme for 
 QKD, it is important to show that the size of the test 
 sample can be a very small fraction of the total number of 
 transmitted photons. 
 
 We shall present some basic constraints here. 
 As will be shown in Section~\ref{s:proof}, these 
 basic constraints turn out to the most important ones. 
 We see from Remark~2 that, if one limits the 
 eavesdropper's information, $I_{eve}$, to less than 
 a small fixed amount, then, as the length, $k$, of the 
 key increases, the allowed infidelity in Theorem~2, $\delta$, of the state must 
 decrease at least as $O(1/k)$. Suppose $m_1$ and $m_2$ signals 
 are tested for the two different bases respectively, it is quite clear 
 that $\delta$ is at least 
 $e^{ O(m_i)}$. This leads to a constraint that 
 $m_i $ is at least $\Omega(\log k)$.\footnote{Notice that 
 this constraint is {\it weaker} than the usual constraint of 
 $m_i = \Omega (N)$ imposed by various other 
 proofs\cite{mayersqkd,biham}. In the next section, we will see 
 that it is, indeed, unnecessary to impose $m_i  = \Omega (N)$.} 
 Suppose $N$ photons are transmitted and Alice sends 
 photons along the rectilinear and diagonal bases with 
 probabilities, $p$ and $1-p$ respectively. 
 Then, the average number of particles available for 
 testing along the rectilinear basis is only $N p^2$. 
 Imposing that $m_i$ is no more than order $N p^2$, 
 we obtain $ N p^2 = \Omega (\log k)$. 
 
\section{Details of Proof of security of efficient QKD} 
\label{s:proof} 
 We will now tighten some of the loose ends in the 
 proof of unconditional 
 security of our efficient QKD protocol, Protocol~E. 
 
\subsection{Using only one basis to generate the raw key} 
 Recall that, in a refined data analysis, 
 Alice and Bob separate data from 
 the two bases into two sets and compute the error rates in the 
 two sets 
 individually. This gives them individual estimates on 
 the bit error rates, $e_1$ and $e_2$, of the $Z$ and $X$ bases 
 respectively. Alice and Bob 
 demand that both error rates must be sufficiently small, say, 
\begin{equation} 
 0 \leq e_1 , e_2 < e_{max}  - \delta_e , 
\label{eq:channel2} 
\end{equation} 
 where $\delta_e$ is some small positive parameter. 
 From the work of Shor-Preskill, $e_{max}$ is about 11\%. 
 
 We would like to derive the relationship 
 between the error rates in the two bases ($X$ and $Z$) in biased 
 BB84 and 
 the bit-flip and phase error rates in the underlying entanglement 
 purification protocol (EPP). Actually, this depends on 
 how the key is generated. In our protocol~E, 
 the raw key is generated only 
 from polarization data in the $X$-basis (diagonal basis), 
 the bit-flip error rate is simply the $X$-basis 
 bit error rate and the phase error rate is simply the 
 $Z$-basis (rectilinear basis) bit error rate. 
 Therefore, no non-trivial conversion between the 
 error rates of the two bases and the bit-flip and 
 phase error rates needs to be performed. 
 This greatly simplifies our analysis without compromising 
 the efficiency nor security of the scheme. 
 
 Therefore, we have: 
\begin{equation} 
 0 \leq e^{phase}_{sample} , e^{bit-flip}_{sample} < 
 e_{max}  - \delta_e , 
\label{eq:channel3} 
\end{equation} 
 where $\delta_e$ is some small positive parameter 
 and $e_{max}$ is about 11\%.

\subsection{Using classical random sampling theory to establish 
 confidence levels} 
 A main point of Shor-Preskill's proof is that the 
 bit-flip and phase error rates of the random sample provide 
 good estimates of the population bit-flip and phase 
 error rates. Indeed, our refined data analysis, 
 as presented in \cite{patent} and earlier version of the 
 current paper, 
 has been employed by Gottesman and Preskill \cite{squeeze} in 
 their recapitulation of Shor and Preskill's proof. 
 Gottesman and Preskill assumed that Alice and Bob generate the 
 key by always measuring along the $Z$-axis. 
 We remark that the problem of establishing confidence levels of 
 the population from the data provided 
 by a random sample is strictly a problem in {\it classical} 
 random sampling theory because the relevant operators 
 all commute with each other. See subsection~\ref{subsec:quality} for 
 details. 
 
 It should be apparent that Gottesman-Preskill's reformulation of 
 Shor-Preskill's proof and its accompanying analysis of 
 classical statistics carry over to our efficient QKD 
 scheme, provided that we employ the 
 prescribed refined data analysis. 
 
 Let us now give more details of the argument that the sample 
 (bit-flip and phase) error rates provide good estimates of 
 the population (bit-flip and phase) error rates. It is simpler to 
 take the 
 limit of $N$ goes to infinity. In this case, the classical 
 de Finetti's representation theorem applies \cite{caves}. 
 The de Finetti's theorem states that the number, $r_1$, of phase 
 errors in the test sample of $m_1$ photons is given by: 
\begin{equation} 
 p (r_1, m_1) = { m_1 \choose r_1 }  \int_0^1 z^{r_1} ( 1-z)^{m_1 - 
 r_1} 
 P^1_{ \infty } (z) dz 
\end{equation} 
 for some `probability of probabilities' (i.e., a non-negative 
 function, $ P^1_{\infty}$). Physically, it means that one 
 can imagine that each photon is generated by some unknown 
 independent, identical distribution that is chosen with a 
 probability, $P^1_{\infty} (z) $. 
 
 Similarly, for the bit-flip errors, 
 its number, $r_2$, in the test sample of $m_2$ photons is given by: 
\begin{equation} 
 p (r_2, m_2) = { m_2 \choose r_2}  \int_0^1 z^{r_2} ( 1-z)^{m_2 - 
 r_2} 
 P^2_{\infty} (z) dz 
\end{equation} 
 for some `probability of probabilities', $P^2_{\infty} (z) dz $. 
 
 We are interested in the case of a finite population size, $N$. 
 Fortunately, a similar expression still 
 exists\cite{lee,renes,jaynes} and it can be written in terms of 
 hypergeometric functions: 
\begin{equation} 
 p (r_2, m_2) = \sum_{ n=r_2}^{ N - m_2 + r_2 } 
 [ C(m_2, r_2) C(N-m_2, n -r_2)/ C(N,n) ] P (n, N) 
\end{equation} 
 where $C(a, b)$ is the number of ways of choosing $b$ 
 objects from $a$ objects and $P (n, M)$ is the `probability of 
 probabilities'. 
 
 An upper bound, which will be sufficient for our purposes, can be 
 found in the following Lemma. 
 
 {\bf Lemma~1}. 
 Suppose one is given a population of $n_{\rm total}$ balls out of 
 which 
 $p n_{\rm total}$ of them are white and the rest are black. One 
 then picks 
 $n_{\rm test}$ balls randomly and uniformly from this population 
\emph{without replacement}. Then, the probability of getting at 
 most 
 $\left\lfloor \lambda n_{\rm test} \right\rfloor$ white balls, 
 $Pr_{\rm wr} (X<\left\lfloor \lambda n_{\rm test} \right\rfloor)$, 
 satisfies the inequality 
\begin{eqnarray} 
 Pr_{\rm wr} (X\leq \left\lfloor \lambda n_{\rm test} 
\right\rfloor) < 
 2^{-n_{\rm test} \{ A(\lambda,p) - n_{\rm test} / [(n_{\rm total} 
 - n_{\rm test})\ln 2] \} } 
\label{E:Pr_inequality} 
\end{eqnarray} 
 provided that $n_{\rm test} > 1$ and $0 \leq \lambda < p$, where 
\begin{equation} 
 A(\lambda,p) = -H(\lambda) - \lambda \log_2 p - (1-\lambda) 
\log_2 (1-p) 
\label{E:A_Def} 
\end{equation} 
 with $H(\lambda) \equiv -\lambda \log_2 \lambda - (1-\lambda) 
\log_2 
 (1-\lambda)$ being the well-known binary entropy function. 
 
 Furthermore, $A(\lambda,p) \geq 0$ whenever $0\leq \lambda \leq p 
 < 1$ and the 
 equality holds if and only if $\lambda = p$.

 {\it Proof}: 
 We denote the probability of getting exactly $j$ white balls by 
 $Pr_{\rm wr} 
 (X=j)$. Clearly, 
\begin{eqnarray} 
 & & Pr_{\rm wr} (X=j) \nonumber \\ 
 & = & \frac{ \left( \!\!\!\begin{array}{c} n_{\rm test} \\ j 
\end{array} 
\!\!\!\right) \!(p n_{\rm total}\!-\!j\!+\!1)_j \ ([1-p] 
 n_{\rm total}\!-\!n_{\rm test}\!+\!j\!+\!1)_{n_{\rm test}-j}}{ 
 (n_{\rm total}\!-\!n_{\rm test}\!+\!1)_{n_{\rm test}}} ~, 
\label{E:Pr_wr_def} 
\end{eqnarray} 
 where $(x)_j \equiv x (x+1) (x+2) \cdots (x+j-1)$. 
 Eq.~(\ref{E:Pr_wr_def}) is called the hypergeometric distribution 
 whose properties have been studied in great detail. In particular, 
 
 Sr\'{o}dka showed that \cite{Probability_Bound} 
\begin{eqnarray} 
 Pr_{\rm wr} (X=j) & < & \left( \!\!\!\begin{array}{c} n_{\rm 
 test} \\ j 
\end{array} \!\!\!\right) p^j (1-p)^{n_{\rm test}-j} \left( 1 - 
\frac{n_{\rm test}}{n_{\rm total}} \right)^{-n_{\rm test}} \times \nonumber \\ 
 & & ~~~~\left[ 1 + \frac{6 n_{\rm test}^2 + 
 6 n_{\rm test} - 1}{12n_{\rm total}} \right]^{-1} \nonumber \\ 
 & < & \left( \!\!\!\begin{array}{c} n_{\rm test} \\ j \end{array}
\!\!\!\right) p^j (1-p)^{n_{\rm test}-j} \left( 1 - 
\frac{n_{\rm test}}{n_{\rm total}} \right)^{-n_{\rm test}} 
\label{E:Pr_wr_bound} 
\end{eqnarray} 
 whenever $n_{\rm test} > 1$. 
 
 Consequently, 
\begin{eqnarray} 
 & & Pr_{\rm wr} (X\leq \left\lfloor \lambda n_{\rm test} 
\right\rfloor) 
\label{e:lemma1a} \\ 
 & < & \left( 1 - \frac{n_{\rm test}}{n_{\rm total}} 
\right)^{-n_{\rm test}} 
\,\sum_{j=0}^{\left\lfloor \lambda n_{\rm test} \right\rfloor} 
\left( 
\!\!\!\begin{array}{c} n_{\rm test} \\ j \end{array} 
\!\!\!\right) p^j 
 (1-p)^{n_{\rm test}-j} 
\label{e:lemma1b} \\ 
 & < & \left( 1 - \frac{n_{\rm test}}{n_{\rm total}} 
\right)^{-n_{\rm test}} 
\,2^{n_{\rm test} [H(\lambda) + \lambda \log_2 p + (1-\lambda) 
\log_2 (1-p)]} 
\label{e:lemma1c} \\ 
 & < & 
 2^{-n_{\rm test} \{ -H(\lambda) - \lambda \log_2 p - (1-\lambda) 
\log_2 
 (1-p) - n_{\rm test} / [(n_{\rm total} - n_{\rm test}) \ln 2]\}} 
\label{e:lemma1d} 
\end{eqnarray} 
 whenever $0\leq \lambda < p$. 
 Note that we have used the inequality in \cite{Coding} 
 to obtain Eq.~(\ref{e:lemma1c}) and 
 the inequality $- { x \over 1 -x} \leq \ln ( 1-x)  \leq -x \leq 0$ 
 to obtain Eq.~(\ref{e:lemma1d}) respectively. Hence, 
 Eq.~(\ref{E:Pr_inequality}) holds. 
 
 Finally we want to show that $A(\lambda,p) \geq 0$ whenever $0\leq 
\lambda \leq p < 
 1$; and the equality holds if and only if $\lambda = p$. This fact 
 follows 
 directly from the observations that $A(\lambda,\lambda) = 0$, 
 $\partial A / 
\partial p \geq 0$ whenever $0\leq \lambda \leq p < 1$ and the 
 equality holds if and only if $\lambda = p$. \hfill Q.E.D. 
 
 Note that Lemma~1 gives a precise bound, 
 not just an approximation. 
 The upshot of Lemma~1 is that the probability that 
 the sample mean deviates from the population mean by any 
 arbitrary but fixed non-zero 
 amount can be shown to be 
 {\it exponentially} small in $n_{test}$, 
 as discussed in subsection~\ref{ss:practical}. 
 In effect, Lemma~1 gives the conditional probability, 
 $\varepsilon_1$, that the signal 
 quality check stage is passed, given that more than $t \equiv 
\left\lfloor (d-1)/2 \right\rfloor$ out of the $n$ pairs of shared 
 entangled particles between Alice and Bob are in error. 
 We will choose $n_{test} = m_1 =m_2$ in our Protocol~E. 
 
\subsection{Bounding fidelity} 
 
 Given any eavesdropping strategy that will pass the verification 
 test with a probability, $\varepsilon_2$, 
 it is important to obtain a bound on the fidelity of the 
 recovered state as $k$ EPR pairs, after quantum error correction 
 and quantum privacy amplification. 
 We have the following Theorem.

 {\bf Theorem~3. (Adapted from \cite{qkd})} Suppose Alice and 
 Bob perform a stabilizer-based EPP-based QKD and, for the 
 verification test, 
 randomly sample 
 along at least two of the three bases, 
 $X$ and $Y$ and $Z$ and compute their error rates. 
 Suppose further 
 that the CSS code used in the signal privacy amplification 
 stage acts on $n$ imperfect pairs of qubits to distill 
 out $k$ pairs of qubits. Given any 
 fixed but arbitrary eavesdropping strategy by Eve, 
 define the following probabilities: 
\begin{equation} 
 p = P( {\rm EPP~succeeds}) , 
\end{equation} 
\begin{equation} 
\varepsilon_1 = P({\rm verification~passed}~| 
 {\rm EPP~fails}), 
\end{equation} 
 and 
\begin{equation} 
\bar{\varepsilon}_1 = P({\rm verification~failed}~| 
 {\rm EPP~succeeds}), 
\end{equation} 
 (In statistics 
 language, $\varepsilon_1$ and $\bar{\varepsilon}_1$ are the type~I 
 and~II 
 errors respectively.) 
 Then, for any Eve's cheating strategy whose probability of passing 
 the 
 verification test is greater 
 than $\varepsilon_2$, the fidelity of the remaining untested 
 shared entangled 
 state immediately after the quantum privacy amplification is 
 greater than 
 $1 - \varepsilon_1 / \varepsilon_2 $.

 {\it Proof}: 
 From Theorem~1 and Proposition~1, one can, indeed, 
 apply classical arguments to 
 the problem by assigning classical probabilities to the 
 $N$-Bell-basis 
 states. Given any fixed but arbitrary eavesdropping strategy, 
 the fidelity of the remaining untested 
 entangled state is given by: 
\begin{eqnarray} 
 F &\geq& { P( {\rm verification~passed~and~EPP~succeeds}) \over 
 P ( {\rm verification~passed}) } \nonumber \\ 
 ~&=& { P( {\rm EPP~succeeds} ) P ({\rm 
 verification~passed}~|{\rm EPP~succeeds}) \over 
 P( {\rm EPP~succeeds} ) P ({\rm 
 verification~passed}~|{\rm EPP~succeeds}) + P( {\rm EPP~fails} ) P 
 ({\rm 
 verification~passed}~|{\rm EPP~fails})}  \nonumber \\ 
 ~&=& { P( {\rm EPP~succeeds} ) P ({\rm 
 verification~passed}~|{\rm EPP~succeeds}) \over 
 P( {\rm EPP~succeeds} ) P ({\rm 
 verification~passed}~|{\rm EPP~succeeds}) + P( {\rm EPP~fails} ) P 
 ({\rm 
 verification~passed}~|{\rm EPP~fails}) }  \nonumber \\ 
 ~&=& { p ( 1 -  \bar{\varepsilon}_1 ) \over 
 p ( 1 -  \bar{\varepsilon}_1 ) + ( 1- p) \varepsilon_1 } 
\nonumber \\
 ~&\geq & 1 - \frac{\varepsilon_1}{p (1-\bar{\varepsilon}_1) + 
 ( 1- p) \varepsilon_1 } ~, 
\label{E:F1} 
\end{eqnarray} 
 
 Now, for any Eve's cheating strategy whose probability of 
 passing the verification test is greater 
 than $\varepsilon_2$, we have $ p (1-\bar{\varepsilon}_1) + 
 ( 1- p) \varepsilon_1 > \varepsilon_2$ and, hence, from 
 Eq.~(\ref{E:F1}), 
\begin{equation} 
 F > 1 - \frac{\varepsilon_1}{\varepsilon_2} ~. \label{E:F2} 
\end{equation} 
 This completes the proof of Theorem~3. \hfill Q.E.D.

\subsection{Summary of the proof} 
\label{ss:summary} 
 We will now put all the pieces together and show that 
 a rigorous proof of security is possible with the number of 
 test particles, $m_1 =m_2 =n_{test}$, scaling 
 logarithmically with the length $k$ of the final key. 
 Consequently, the bias in an efficient BB84 scheme can be 
 chosen such that $N ( p^2 - \delta') = n_{test}$ for a small 
 $\delta$. In other words, $p = O ( \sqrt{(\log k) / N })$, which 
 goes to zero as $N$ goes to infinity. 
 
 Given a signal quality check that involves only $n_{test}$ 
 photons, from Lemma~1, we see that 
 the conditional probability, $ \varepsilon_1$, 
 that the signal 
 quality check stage is passed, given that more than $t \equiv 
\left\lfloor (d-1)/2 \right\rfloor$ out of the $n$ pairs of shared 
 entangled particles between Alice and Bob are in error is 
 exponentially small in $n_{test}$. i.e., 
\begin{equation} 
\varepsilon_1 = O (2^{- n_{test} \alpha}),
\label{e:epsilon1} 
\end{equation} 
for some positive constant $\alpha$.
 
 Let Alice and Bob pick a security parameter, 
\begin{equation} 
\varepsilon_2 = 2^{-u}, 
\label{e:defineu} 
\end{equation} 
 and consider only eavesdropping strategies that will pass 
 the signal quality check with a probability at least 
 $ \varepsilon_2$. We require that 
\begin{equation} 
\varepsilon = { \varepsilon_1 \over \varepsilon_2} 
\ll 1. 
\label{e:approximation} 
\end{equation} 
 
 Recall from Theorem~3 that for any eavesdropping strategy 
 that will pass the signal quality check test with a probability 
 at least $\varepsilon_2$, has its fidelity bounded by 
 $ 1 - \varepsilon$. i.e., 
\begin{equation} 
 F \geq 1 -  \varepsilon. 
\end{equation} 
 
 Now, from Theorem~2, the eavesdropper's mutual information 
 with the final key is bounded by 
\begin{equation} 
 I_{eve}^{Bound} = \varepsilon ( 2 k + \log_2( 1/ \varepsilon ) +  { 1 \over 
\log_e 2}) . 
\label{e:bound} 
\end{equation} 
 
 Consider a fixed but arbitrary value of $I_{eve}^{Bound}$, the constraint on 
 the eavesdropper's mutual information on the final key: 
 i.e., 
\begin{equation} 
 I_{eve}^{Bound} = 2^{- s}, 
\label{e:securitys} 
\end{equation} 
 where $s$ is a positive security parameter. 
 In the large $k$ limit, Eq.~(\ref{e:bound}) implies 
 that 
\begin{equation} 
\varepsilon = O ( 2^{-s}/k). 
\label{e:orderk} 
\end{equation} 
 Substituting Eq.~(\ref{e:approximation}) into 
 Eq.~(\ref{e:orderk}), we see that 
\begin{equation} 
 { k \varepsilon_1 \over  2^{-s} \varepsilon_2} = O (1). 
\label{e:order1} 
\end{equation} 
 Substituting Eqs.~(\ref{e:epsilon1}) and~(\ref{e:defineu}) into 
 Eq.~(\ref{e:order1}), we find that 
\begin{equation} 
 { k 2^{- n_{test} \alpha} \over  2^{-(u + s) } } = O (1). 
\label{e:order2} 
\end{equation} 
 
 Now, for fixed but arbitrary values of the security parameters, 
 $s$ and $u$, we see that, in fact, the number of test photons, $n_{test}$, 
 is required to scale only as $ O (\log k)$, i.e., the logarithm of 
 the final key length. Consequently, the only constraint on the 
 bias $p$ is that there are enough photons for performing the 
 verification test. This gives rise to the requirement that 
 $N ( p^2 - \delta') = n_{test} = O (\log k)$, i.e., 
\begin{equation} 
 p = O ( \sqrt{(\log k) / N }). 
\end{equation} 
 
 This completes our proof of security of Protocol~E, 
 an efficient QKD scheme. 
 We remark that the error correction and privacy amplification 
 procedure in Protocol~E are exactly the same as in Shor-Preskill's 
 proof. 
 
 As a side remark, if one insists that the eavesdropper's 
 information is exponentially small in $N$, then one can 
 take $s = c N$, for some positive constant, $c$. 
 From Eq.~(\ref{e:order2}), this will require $n_{test}$ 
 to be proportional to $N$. A number of earlier papers make 
 such an assumption. However, in this paper, we note that 
 this requirement can be relaxed. For instance, it is consistent to 
 pick $s = c N^{a'}$ where $ 0 \leq a' \leq 1$. 
 In this more general case, we have from Eq.~(\ref{e:order2}) 
 that asymptotically $\alpha n_{test} \sim c N^{a'}$. 
 Consequently, 
\begin{eqnarray} 
\alpha N p^2 &\geq& \alpha n_{test} \sim c N^{a'} \nonumber \\ 
 p^2 & = &\Omega  ( { c N^{a' -1} \over \alpha } )  . 
\label{e:a'} 
\end{eqnarray} 
 {}From Eq.~(\ref{e:a'}), it is clear that 
 for all values of $a' \in [0,1]$, the probability 
 $p$ can be chosen to be arbitrarily 
 small, but non-zero. This completes our analysis for 
 the security of an efficient QKD scheme where each of 
 Alice and Bob picks the two polarization bases with probabilities 
 $p$ and 
 $1-p$. 
 
\section{Concluding Remarks} 
\label{sec:conclusion} 
 In this paper, we presented a new quantum key distribution scheme 
 and 
 proved its unconditional security against the most general attacks 
 allowed by quantum mechanics. 
 
 In BB84, each of Alice and Bob chooses between the two bases 
 (rectilinear and 
 diagonal) with equal probability. Consequently, Bob's measurement 
 basis differs from that of Alice's half of the time. For this 
 reason, 
 half of the polarization data are useless and are thus thrown away 
 immediately. 
 We have presented a simple modification that can essentially double 
 the efficiency of BB84. There are two important ingredients in this 
 modification. 
 The first ingredient is for each of Alice and Bob to assign 
 significantly 
 different probabilities (say $\epsilon$ and $1 - \epsilon $ 
 respectively where 
 $\epsilon$ is small but non-zero) to the two polarization bases 
 (rectilinear 
 and diagonal respectively). Consequently, they are much more likely 
 to 
 use the same basis. This decisively enhances efficiency. 
 
 However, an eavesdropper may try to break such a scheme by 
 eavesdropping mainly along the predominant basis. 
 To make the scheme secure against such a biased eavesdropping 
 attack, 
 it is crucial to have the second 
 ingredient --- a refined error analysis --- in place. 
 The idea is the following. Instead of lumping all the accepted 
 polarization 
 data into one set and computing a {\em single\/} error rate (as in 
 BB84), we 
 divide up the data into various subsets according to the actual 
 polarization 
 bases used by Alice and Bob. In particular, the {\em two\/} error 
 rates for 
 the cases 1) when both Alice and Bob use the rectilinear basis and 
 2) when both Alice and Bob use the diagonal basis, are computed 
 separately. 
 It is only when both error rates are small that they accept the 
 security of 
 the transmission. 
 
 We then prove the security of efficient QKD scheme, not only against 
 the 
 specific attack mentioned above, but also against the 
 most general attacks allowed by the laws of quantum 
 mechanics. In other words, our new scheme is {\em unconditionally\/} 
 secure. 
 Moreover, just like the standard BB84 scheme, our protocol can be 
 implemented 
 without a quantum computer. 
 The maximal tolerable bit error rate is 11\%, the same as in 
 Shor and Preskill's proof. 
 If we allow Eve to get a fixed but arbitrarily small amount of 
 information 
 on the final key, then the number of test particles, $n_{test}$, 
 is required only to scale 
 logarithmically with the length $k$ of the final key. 
 Consequently, the bias in an efficient BB84 scheme can be 
 chosen such that $N ( p^2 - \delta') = n_{test}$ for a small 
 $\delta$ and where $N$ is the total number of 
 photons transmitted. In other words, $p = O ( \sqrt{(\log k) / N 
 })$, 
 which goes to zero as $N$ goes to infinity. 
 More generally, 
 suppose we pick the security parameter to be $s$ 
 (for an eavesdropper's information $I_{eve} \leq 2^{-s}$) such that 
 $s = c N^{a'}$ where $ 0 \leq a' \leq 1$. We find that this 
 can be achieved by testing $n_{test}$ random photons where 
 $\alpha n_{test} \sim c N^{a'}$. Furthermore, each of Alice and 
 Bob may pick the two polarization 
 bases with probabilities $p$ and $1-p$ such that 
 $p^2 = \Omega ( { c N^{a' -1} \over \alpha }) $. Therefore, 
 $p$ can, indeed, be made arbitrarily small but non-zero. 
 
 This is the first time that a single-particle 
 quantum key distribution scheme has been proven to be secure 
 without relying on a symmetry argument --- that the two bases are 
 chosen 
 randomly and uniformly. 
 Our proof is a generalization of 
 Shor and Preskill's proof \cite{shorpre} of security of BB84, 
 a proof that in turn 
 built on earlier proofs by Lo and Chau \cite{qkd} and 
 also by Mayers \cite{mayersqkd}. 
 
 We remark that our idea of efficient schemes of quantum key 
 distribution 
 applies also to other schemes such as Biham, Huttner and Mor's 
 scheme \cite{Mor} which is 
 based on quantum memories. 
 Our idea also applies the six-state scheme \cite{bruss}, 
 which has been shown rigorously to tolerate a higher error rate of 
 up to 12.7\% \cite{six}. 
 
 As a side remark, Alice and Bob may use different biases 
 in their choices of probabilities. In other words, 
 our idea still works if Alice chooses between 
 the two bases with probabilities $\epsilon$ and  $1- \epsilon$ 
 and Bob chooses with probabilities $\epsilon'$ and  $1- \epsilon'$ 
 where $\epsilon \not= \epsilon'$. 
 
 We thank Gilles Brassard for many helpful discussions and 
 suggestions. 
 Enlightening discussions with Daniel Gottesman, Debbie Leung, 
 Norbert 
 L\"{u}tkenhaus, John 
 Preskill and Peter Shor on various proofs of 
 the unconditional security of QKD schemes are also 
 gratefully acknowledged. We also thank Peter Shor for suggesting to 
 us the possibility that Shor and Preskill's proof may be generalized 
 to prove 
 the unconditional security of efficient quantum key distribution 
 scheme. We gratefully acknowledge enlightening discussions with 
 Chris Fuchs, Stephen Lee, and Joe Renes about 
 statistical analysis. We thank P. L. H. Yu for his useful 
 discussions on 
 hypergeometric function and pointing out some references to us. 
 We also thank anonymous referees for their many comments, 
 which are very useful for improving the presentation of the 
 current paper. 
 HFC is supported by the RGC grant HKU~7143/99P of 
 the Hong Kong SAR Government. Parts of this work were done while M. 
 Ardehali 
 was with NEC Japan and while H.-K. Lo was at the Institute for 
 Advanced Study, Princeton, NJ, Hewlett-Packard Laboratory, 
 Bristol, UK and MagiQ Technologies, Inc., New York.
 
\par\bigskip\noindent 
 {\it Notes Added}: An entanglement-based scheme with an 
 efficiency greater than 
 $50\%$ has also been discussed in a recent preprint by two of us (H.-K. Lo  and H.F. Chau) \cite{qkd}. 
 Recent proofs of the unconditional security of 
 various QKD schemes have been provided 
 by H. Inamori \cite{hitoshi1,hitoshi2}, 
 H. Aschauer and H. J. Briegel \cite{hans} and by 
 D. Gottesman and J. Preskill \cite{squeeze}. 
 Recently, it has been shown \cite{two} by D. Gottesman and 
 one of us (H.-K. Lo) 
 that two-way classical communications can be used to increase 
 substantially the maximal tolerable bit error rate in BB84 and 
 the six-state scheme. The result presented in the current
paper can be combined with \cite{two} to obtain, for example, an efficient
BB84 scheme that can tolerate a substantially higher bit error
rate (say, 18.9 percent) than in Shor-Preskill's proof.
It has been shown in a recent preprint \cite{imperfect}
that even imperfect devices can provide perfect security in QKD within the
entanglement purification approach employed in the present paper.
Finally, a proof of the unconditional security of another well-known
QKD scheme, B92 scheme published by Bennett in 1992 \cite{b92}, has
recently been presented \cite{proofb92}.

\end{document}